\documentclass[aps,prb,twocolumn,secnumarabic,amsmath,amssymb,superscriptaddress]{revtex4-1}
\usepackage{dcolumn}
\usepackage{bm}

\usepackage{color} 
\usepackage{ulem}

\usepackage{amsmath}
\usepackage{graphicx}
\usepackage{float}
\usepackage{subfig}

\usepackage{color}
\usepackage[colorlinks,bookmarks=false,citecolor=blue,linkcolor=red,urlcolor=blue]{hyperref}

\definecolor{darkred}{rgb}{0.7,0.0,0.0}

\definecolor{darkblue}{rgb}{0,0.02,0.45}

\definecolor{darkgreen}{rgb}{0.02,0.45,0.0}

\definecolor{violet}{rgb}{0.8,0.2,0.6}

\providecommand{\U}[1]{\protect\rule{.1in}{.1in}}

\begin{document}

\title{Majorana orthogonal transformation and Majorana zero modes in free fermionic systems}

\author{Jianlong Fu}
\affiliation{Department of Physics, The University of Hong Kong, Pokfulam Road, Hong Kong, China}

\begin{abstract}
 We study free fermionic models that host Majorana zero modes using the Majorana orthogonal transformation, which is a type of transformation between different fermionic models under Majorana representation of complex fermions. Using Majorana orthogonal transformation, a U(1) topological gauge theory for the doubled $p_{x}+ip_{y}$ topological superconductor is obtained; the vortex Majorana zero modes and the degeneracy splitting of multiple vortices are studied using field theoretical method. For lattice Majorana hopping models, we perform real-space analysis on the Majorana zero modes. In one dimension, the decoupled Su-Schrieffer-Heeger model and the Kitaev chain are discussed as examples and building blocks for composite models. In two dimensions a simple lattice model realizing the $p_{x}+ip_{y}$ superconductor is introduced, and its defect Majorana zero mode is written down explicitly. We introduce a systematic way to obtain models hosting Majorana zero modes in which composite models are constructed from two independent Majorana hopping models by Majorana orthogonal transformations. Three one-dimensional models are proposed and discussed as examples.
\end{abstract}

\maketitle
\section{Introduction}

Topological phases of matter has attracted a lot of attention in the past decades in condensed matter physics \cite{fradkinbook,schnyder08,chiu16,kitaev09,Read00,kane051,kane052,qi11,hasan10,freedman04,tknn82,laughlin81,halperin82,Kitaev2003,Kitaev2006}. One of the physical characteristics of topological phases is the existence of zero modes (or gapless modes) associated with edges or defects of the system \cite{fradkinbook,halperin82,schnyder08,chiu16,Read00,Volovikbook,girvinbook}. For a many-body system, a zero mode is a {\it degeneracy} in the many-particle spectrum which corresponds to a zero-energy state of the single-particle Hamiltonian. The mathematical indicators of the topological phases are different kinds of bulk {\it topological numbers} \cite{chiu16,tknn82,kane051,kane052}, and the connection between a nonzero bulk topological number and the appearance of edge or defect zero modes is part of the {\it bulk boundary or bulk defect correspondence}\cite{mooreread91,chiu16,Qi2006,teo2010}. Among different types of zero modes, the edge gapless modes in topological models of more than one spatial dimension are important because they are responsible for the transport properties of the system \cite{halperin82}; on the other hand, the point-defect zero modes are interesting since they can sometimes be treated as {\it particles} which can move around in the topological systems. In two dimensions (2D) specifically, these particles can have {\it exotic statistics} instead of just bosonic or fermionic \cite{wilczek82}; such exotic particles are called {\it anyons} and they have many potential applications in quantum computation \cite{nayak08,Kitaev2003,Kitaev2006,Sarmareview}. In this paper, we are interested in gapped {\it free} fermionic systems, including insulators and superconductor Bogoliubov-de Gennes (BdG) systems. In these systems, a complete {\it ten-fold way} classification is achieved based on symmetries and topological numbers \cite{schnyder08,chiu16,altland1997,teo2010}. As opposed to interacting systems in 2D, which can host many types of anyonic statistics \cite{Laughlin83,nayak08,Kitaev2006}, in {\it free} fermionic systems the possible particle types are limited. For defect zero modes in free systems, the only exotic type is {\it Majorana fermion} \cite{Wilczekreview,Elliott15}. To be specific a Majorana fermion is a fermionic particle that is its own antiparticle, namely the Majorana operator $\eta_{i}$ satisfy $\eta_{i}^{\dagger}=\eta_{i}$; for two Majorana fermions $\eta_{i}$ and $\eta_{j}$, we have the anticommutation relation $\{\eta_{i},\eta_{j}\}=2\delta_{ij}$. In condensed matter physics, a single spinless complex fermion $c_{i}$ can be decoupled into two Majorana fermions $\eta_{i}^{\alpha}$, $\eta_{i}^{\beta}$: $c_{i}^{\dagger}=\frac{1}{2}(\eta_{i}^{\alpha}+i\eta_{i}^{\beta})$; conversely two Majorana fermions can be paired up to form a complex fermion; such relation can be intepretted as {\it Majorana representation of complex fermion}. In ideal situation, a {\it Majorana zero mode} of a topological system is a Majorana operator that commutes with the Hamiltonian \cite{Sarmareview}. In 2D systems, Majorana zero modes have non-Abelian statistics when braiding among themselves \cite{nayak08,fradkinbook,Kitaev2006, Ivanov01}. 

Due to the exotic nature and applications there have been a lot of attention on Majorana zero modes in topological systems \cite{alicea2012,beenakker2013,beenakker2015,stanescu2013,Sarmareview,Elliott15}. Besides strongly correlated systems such as the $\frac{5}{2}$-quantum Hall state \cite{Willett87,Moore91,Greiter92} and the Kitaev-type spin models \cite{Kitaev2006,fu20191}, a few important {\it free} models that host defect Majorana zero modes have been proposed, these include the 2D $p_{x}+ip_{y}$ topological superconductor \cite{fradkinbook,Read00,volovik99} and the 1D $p$-wave superconducting Kitaev chain \cite{Kitaev2001}. Specifically the full vortex defects in the 2D $p_{x}+ip_{y}$ superconductor of spinless fermions support Majorana zero modes, which has been argued theoretically using different approaches \cite{Read00,roy10,gurarie07,nishida2010,tewari07}. Experimentally it has possible realization in $\mathrm{Sr}_{2}\mathrm{RuO}_{4}$ \cite{sarma2006,machenzie2003}. Another way to produce the $p_{x}+ip_{y}$ superconductor effectively is by combining an $s$-wave superconductor with topological insulator \cite{ful08,ful09}, or quantum anomalous Hall system \cite{Qitheory10,Wangtheory15}; experimental results in search of vortex Majorana zero modes have been positive in these systems \cite{Xuexp15,He294}. Theoretically, the 1D Kitaev chain has Majorana edge zero modes; but it is not directly obtainable experimentally since it involves $p$-wave superconducting spinless fermions. Recently a few possible schemes were proposed to realize the Kitaev chain indirectly. One of them is the semiconductor nanowire-superconductor system \cite{sau2010,alicea2010,lutchyn2010,oreg10,Sau10theory}; signals of Majorana zero modes have been observed in such systems \cite{Mourik1003,rokhinson12,Churchill13,Das12,Finck13,Deng1557}. Another way to synthesize the Kitaev chain is by using the ferromagnetic atoms on superconductor surface \cite{Nadjperge13,Choy11,Pientka13,Klinovaja13,Braunecker13,Litheory14}, and these systems also yield positive experimental results \cite{Nadj-Perge602,Ruby15}.       

The aforementioned theoretical studies of Majorana zero modes share similar interests, which include what models can host Majorana zero modes and why Majorana zero modes appear in these models. Traditional approaches of these questions usually focus on complex fermion models and their bulk topological numbers in momentum space. In this paper, we formulate systematically a {\it real-space} method to study Majorana zero modes in various topological systems based on the {\it Majorana representation of complex fermions}. Our approach starts with the fact that all free complex fermionic models can be mapped into {\it Majorana hopping models} under Majorana representations; conversely for a given Majorana hopping model, the Majorana fermions can be paired up in different ways to form different complex hopping models. We argue that this process corresponds to the {\it Majorana orthogonal transformations}, which define {\it dualities} between complex fermion models \cite{cobanera2015}. General Majorana orthogonal transformations include {\it stacking two different models or doubling a single model} and gluing them together by Majorana fermions decoupling and pairing. These transformations between complex fermion models preserve the spectrum and the symmetries of the model, very importantly they also preserve the existence of zero modes. Using Majorana orthogonal transformations, we can achieve {\it new understanding} of the models that host Majorana zero modes. In particular, by {\it doubling} the system, we obtain a U(1) topological gauge theory of massive Dirac fermion for the {\it continuous} $p_{x}+ip_{y}$ topological superconductor, using which we argue the existence of vortex Majorana zero modes and compute the splitting of degeneracy of multiple vortices from a field-theoretical perspective. For {\it lattice} models hosting Majorana zero modes, we introduce the notion of {\it simple models} which include the decoupled 1D Su-Schrieffer-Heeger (SSH) model and the Kitaev chain with real parameters. We also construct a 2D simple Majorana hopping model that realizes the $p_{x}+ip_{y}$ superconductor at low energies, with a defect Majorana zero mode whose wavefunction can be written down directly. Finally we show that Majorana orthogonal transformations can be used to {\it construct new models} hosting Majorana zero modes by stacking and gluing together two independent models. We discuss examples of such construction in 1D by considering two layers of rotated Kitaev chain. The resulting composite models have spinful complex fermions and various types of superconducting pairing which make them useful to overcome the ``fermion doubling" problem \cite{alicea2012} for finding Majorana zero modes in real materials. And we point out that the applicability of these constructions is limitless.

The rest of the paper is organized as follows. In Sec. \ref{secmajoranarothogonal}, we start with a discussion on the mapping between complex hopping models and Majorana hopping models as well as the Majorana orthogonal transformations; we then move on to discuss the definition and properties of the defect zero modes under Majorana orthogonal transformation and we point out another possible origin of Majorana hopping models from a certain type of exactly solvable interacting models. In Sec. \ref{seccontinuous}, we consider the continuous theory of $p_{x}+ip_{y}$ topological superconductor and obtain a topological U(1) gauge theory for the doubled system. We then move on to lattice models in Sec. \ref{secsimplemodels}, in which we consider three simple models, namely the SSH model, the Kitaev chain and a 2D model realizing the $p_{x}+ip_{y}$ topological superconductor. In Sec. \ref{secgeneralization}, we discuss the application of Majorana orthogonal transformation to construct composite and more complex models and we give three examples in 1D. The paper concludes in Sec. \ref{secconclusion} with some discussions for future studies.

\section{Majorana orthogonal transformations and zero modes in free fermionic models} \label{secmajoranarothogonal}

In this section, we start by considering the relationship between a Majorana hopping model and a complex hopping model. Specifically, we look at how a complex model is decoupled into a Majorana hopping model, and we try to understand how Majorana hopping models are transformed into complex models by pairing up Majorana fermions in a certain way. In these processes, a complex model can be decoupled into a Majorana model which is subsequently paired up in a different way to form a different complex fermion model. We study the relationship between these two complex models and introduce the notion of {\it Majorana orthogonal transformation}\cite{cobanera2015}. We then turn to study the properties of the zero modes under Majorana orthogonal transformations. Finally, we point out another origin of the Majorana hopping models from a certain type of exactly solvable strongly correlated models. 

\subsection{From complex hopping model to Majorana hopping model}

Here we study how a general complex fermion hopping model can be decoupled into a Majorana hopping model. We take the general spinless complex fermion model with BCS pairing term
\begin{equation}
\label{generalcomplexhamiltonian}
\mathcal{H}=\sum_{\langle ij\rangle}t_{ij}c_{i}^{\dagger}c_{j}+t_{ij}^{*}c_{j}^{\dagger}c_{i}+\Delta_{ij}c_{i}c_{j}+\Delta_{ij}^{*}c_{j}^{\dagger}c_{i}^{\dagger}.
\end{equation}
Each complex fermion can be decoupled into two Majorana fermions, which we call $\eta_{i}^{\alpha}$ and $\eta_{i}^{\beta}$,
\begin{equation}
c_{i}^{\dagger}=\frac{1}{2}(\eta_{i}^{\alpha}+i\eta_{i}^{\beta}),\qquad c_{i}=\frac{1}{2}(\eta_{i}^{\alpha}-i\eta_{i}^{\beta}).
\end{equation}
Using these decoupling, the Hamiltonian (\ref{generalcomplexhamiltonian}) is transformed into
\begin{eqnarray}
\begin{aligned}
\label{complextomajorana}
\mathcal{H}=\frac{1}{2}\sum_{\langle ij\rangle}&\big(\operatorname{Im}t_{ij}+\operatorname{Im}\Delta_{ij}\big)i\eta_{i}^{\alpha}\eta_{j}^{\alpha}\\+&\big(\operatorname{Re}t_{ij}-\operatorname{Re}\Delta_{ij}\big)i\eta_{i}^{\beta}\eta_{j}^{\alpha}\\-&\big(\operatorname{Re}t_{ij}+\operatorname{Re}\Delta_{ij}\big)i\eta_{i}^{\alpha}\eta_{j}^{\beta}\\+&\big(\operatorname{Im}t_{ij}-\operatorname{Im}\Delta_{ij}\big)i\eta_{i}^{\beta}\eta_{j}^{\beta}.
\end{aligned}
\end{eqnarray}
One can add a chemical potential term to the complex fermion Hamiltonian (\ref{generalcomplexhamiltonian}), such term is transformed as follows,
\begin{equation}
\label{complextomajoranachemical}
\sum_{i}\mu_{i}c_{i}^{\dagger}c_{i}=\sum_{i}\frac{1}{2}\mu_{i}(1-i\eta_{i}^{\alpha}\eta_{i}^{\beta}).
\end{equation}

Under certain conditions, the resulting Majorana hopping model (\ref{complextomajorana}) and (\ref{complextomajoranachemical}) automatically decouple into two independent Majorana hopping models. In particular, there are two such possibilities. First, for $t_{ij}$ and $\Delta_{ij}$ being purely imaginary and $\mu_{i}\equiv 0$, the complex fermion Hamiltonian can decouple into two independent Majorana hopping models for $\{\eta_{i}^{\alpha}\}$ and $\{\eta_{i}^{\beta}\}$ respectively. Second, sometimes the lattice sites can be grouped into two sets $A$ and $B$ with all the Majorana hopping paths $\langle ij\rangle$ connecting one sites belonging to group $A$ and the other belonging to group $B$. In this situation, if $t_{ij}$ and $\Delta_{ij}$ are purely real and $\mu_{i}\equiv 0$, the complex fermion Hamiltonian also decouples into two layers of independent Majorana hopping models. 

\subsection{From Majorana hopping model to complex hopping models}

Given a Majorana hopping model, there are multiple ways to pair up the Majorana fermions and obtain complex models. As pointed out by the previous section, there are two situations. The Majorana hopping model may form a single connected layer in which every lattice point can reach to any other lattice point following the Majorana hopping path. In another situation the original Majorana hopping model can form two independent layers and still corresponds to a single layer complex fermionic model under certain pairing scheme. 

We start by considering single-layer Majorana hopping models. To obtain a complex hopping model, we first pair up sites within the layer. This scheme will be referred to as {\it intralayer pairing}. For each pair of Majorana fermion we use $i$ and $j$ to label its position. Within the pair $i$, we assign superscripts $\alpha$ and $\beta$ to the two Majorana fermions, $\eta_{i}^{\alpha}$ and $\eta_{i}^{\beta}$, the complex fermion is thus defined by $c_{i}^{\dagger}=\frac{1}{2}(\eta_{i}^{\alpha}+i\eta_{i}^{\beta})$. After the intralayer pairing, the most general single layer Majorana hopping model can be written as
\begin{eqnarray}
\begin{aligned}
\label{generalmajorana}
\mathcal{H}=&\sum_{ij}iA_{ij}^{\alpha\alpha}\eta_{i}^{\alpha}\eta_{j}^{\alpha}+iA_{ij}^{\beta\beta}\eta_{i}^{\beta}\eta_{j}^{\beta}+iA_{ij}^{\alpha\beta}\eta_{i}^{\alpha}\eta_{j}^{\beta}+iA_{ij}^{\beta\alpha}\eta_{i}^{\beta}\eta_{j}^{\alpha}\\&-\sum_{i}\frac{i}{2}\mu_{i}\eta_{i}^{\alpha}\eta_{i}^{\beta}.
\end{aligned}
\end{eqnarray}
All the coefficients $A_{ij}$ and $\mu_{i}$ are real numbers. The complex hopping model in the form of (\ref{generalcomplexhamiltonian}) can be obtained by comparing (\ref{generalmajorana}) with (\ref{complextomajorana}) and (\ref{complextomajoranachemical}).  Specifically, the condition for the resulting complex hopping model to be an insulator is $A_{ij}^{\alpha\alpha}=A_{ij}^{\beta\beta}$ and $A_{ij}^{\alpha\beta}=-A_{ij}^{\beta\alpha}$. If this condition is not satisfied, the resulting complex Hamiltonian is then a supercondutor BdG type of Hamiltonian.

We then move on to consider a double-layer Majorana hopping model. By definition, there is no interlayer hopping between Majorana fermions and there is a one-to-one correspondence between the sites of the two layers hence they can be labeled by the same simbols. The general double layer Majorana Hamiltonian can be written as
\begin{equation}
\label{generaldoublelayermajorana}
\mathcal{H}=\mathcal{H}_{A}\oplus\mathcal{H}_{B}=\sum_{\langle ij\rangle}iA_{ij}\eta_{i}\eta_{j}+iB_{ij}\tilde{\eta}_{i}\tilde{\eta}_{j},
\end{equation}
in which we used $\eta_{i}$ and $\tilde{\eta}_{i}$ to denote corresponding Majorana fermions on the two layers respectively. The two layers can be referred to as layer $A$ and layer $B$.  To obtain a complex fermion model, we choose to pair up the Majorana fermions on the same site from the two layers and define complex fermion
\begin{equation}
\label{definitionofcomplex}
c_{i}^{\dagger}=\frac{1}{2}(\eta_{i}+i\tilde{\eta}_{i}),\qquad c_{i}=\frac{1}{2}(\eta_{i}-i\tilde{\eta}_{i}).
\end{equation}
Conversely we have $\eta_{i}=c_{i}+c_{i}^{\dagger},\qquad \tilde{\eta}_{i}=i(c_{i}-c_{i}^{\dagger})$. According to these, the Hamiltonian (\ref{generaldoublelayermajorana}) can then be written as
\begin{equation}
\label{doublelayer}
\mathcal{H}=\sum_{\langle ij\rangle}i(A_{ij}-B_{ij})(c_{i}c_{j}+c_{i}^{\dagger}c_{j}^{\dagger})+i(A_{ij}+B_{ij})(c_{i}^{\dagger}c_{j}+c_{i}c_{j}^{\dagger}).
\end{equation}
Such pairing can thus be referred to as {\it interlayer pairing}. Specifically, if the two layers have identical hopping coefficients, namely $A_{ij}=B_{ij}$ for every bond, then the resulting Hamiltonian is an insulator instead of a superconductor BdG system. 

Some discussion is in order before we move on. First, for a single-layer Majorana hopping model, it is always possible to {\it double the system} by introducing another copy of the model and subsequently treat them as a double layer system. We will discuss this method further later on in this paper. Second, it is important to note that the transformation between complex hopping model and Majorana hopping model works in any spatial dimension. For example, a single-layer system can be defined on a three-dimensional lattice.

\subsection{From complex hopping model to Majorana hopping model back to another complex hopping model: Majorana orthogonal transformation} \label{SecMOT}

Physical systems are systems of complex fermions. Following the procedures discussed in previous sections, starting from any complex free fermion model, we are able to decouple the complex fermion degrees of freedom into Majorana fermions and then pair them up in a different way to form another complex fermion model. The decoupling and pairing can be arbitrary, and this process defines a transformation between complex fermion models, with the Majorana model acting as an intermediate system \cite{cobanera2015}. This transformation involves both particles and holes, thus it is an unitary transformation on the Nambu spinor space of the original complex fermion system. In this section we explore the nature of this transformation.

The first type of the transformation is for single-layer Majorana hopping models, for which the original physical Hamiltonian can be denoted by $\mathcal{H}(c_{i},c_{i}^{\dagger})$, with $c_{i}$ being the original complex degrees of freedom. As the system is a free fermion model, the corresponding many-particle Hamiltonian can be written as \cite{chiu16}
\begin{equation}
\mathcal{H}(c_{i},c_{i}^{\dagger})=\Psi_{i}^{\dagger}H_{ij}\Psi_{j},
\end{equation}
in which $\Psi$ is the Nambu spinor $\Psi=(c_{1}^{\dagger},\cdots,c_{N}^{\dagger},c_{1},\cdots,c_{N})^{T}$, $N$ is the total number of complex fermions in the system. A decoupling into Majorana fermions and then pairing them up in another way correspond to the following process
\begin{equation}
\label{singlelayernambu}
\{c_{i}^{\dagger}\}\rightarrow\{\eta_{i}^{\alpha},\eta_{i}^{\beta}\}\rightarrow\{d_{i}^{\dagger}\},\qquad i=1,2,\cdots,N.
\end{equation}
Here, $d_{i}$ denotes the final complex fermion degree of freedom. Defining $\Phi=(d_{1}^{\dagger},\cdots,d_{N}^{\dagger},d_{1},\cdots,d_{N})^{T}$, we have $\Phi=U\Psi$, $U^{\dagger}U=I$. For the many-particle Hamiltonian we have
\begin{equation}
\mathcal{H}(c_{i},c_{i}^{\dagger})\rightarrow \mathcal{H}(d_{i},d_{i}^{\dagger})=\Phi^{\dagger}\tilde{H}\Phi.
\end{equation}
The corresponding single particle Hamiltonian satisfies $\tilde{H}=UHU^{\dagger}$.

The second type of transformation is for double-layer Majorana system. For a given double-layer Majorana hopping model, one can perform intralayer pairing for the two layers independently, so that the system corresponds to two layers of complex fermion modes which do not talk to each other. One can take this double-layer complex fermion system as the original physical model and their Hamiltonians can be denoted as $\mathcal{H}_{A}(c_{i}^{A},c_{i}^{A\dagger})$ and $\mathcal{H}_{B}(c_{i}^{B},c_{i}^{B\dagger})$. The total Hilbert space is captured by the Nambu spinor $\Psi=(c_{1}^{A\dagger},\cdots,c_{N}^{A\dagger},c_{1}^{A},\cdots,c_{N}^{A},c_{1}^{B\dagger},\cdots,c_{N}^{B\dagger},c_{1}^{B},\cdots,c_{N}^{B})=\Psi^{A}\oplus\Psi^{B}$. The Hamiltonian acting on the total Hilbert space can be written as
\begin{equation}
\label{directadditionoftwohamiltonian}
\mathcal{H}_{A+B}=\mathcal{H}_{A}\oplus\mathcal{H}_{B}=\Psi^{\dagger}H_{A+B}\Psi.
\end{equation}
The single-particle energy eigenvalues and eigenstates of the system are given by the combination of those of the two subsystems; the many-particle eigenstates and eigenvalues can be obtained accordingly. From the double-layer system, the interlayer paring of Majorana fermions corresponds to the following
\begin{eqnarray}
\begin{aligned}
\label{doublelayernambu}
&\{c_{i}^{A\dagger},c_{i}^{B\dagger}\}\rightarrow\{\eta_{i,A}^{\alpha},\eta_{i,A}^{\beta},\eta_{i,B}^{\alpha},\eta_{i,B}^{\beta}\}\rightarrow \{d_{j}^{\dagger}\},\\ &i=1,\cdots,N,\qquad j=1,\cdots,2N.
\end{aligned}
\end{eqnarray}
Here we use $d_{j}$ to denote the final complex fermion degrees of freedom. Defining $\Phi=(d_{1}^{\dagger},\cdots,d_{2N}^{\dagger},d_{1},\cdots,d_{2N})^{T}$, the transformation (\ref{doublelayernambu}) is then captured by the unitary transformation $\Phi=U\Psi$. Furthermore, we have the Hamiltonian transforms as
\begin{equation}
\mathcal{H}_{A+B}\rightarrow\mathcal{H}(d_{i},d_{i}^{\dagger})=\Phi^{\dagger}\tilde{H}_{A+B}\Phi,
\end{equation}
in which single-particle Hamiltonian $\tilde{H}_{A+B}=UH_{A+B}U^{\dagger}$. In the second type of transformation, by double-layer pairing of Majorana fermions, two layers of independent complex fermion models are {\it added} together into another complex fermion model. In general, this process can be applied to any number of layers. 

From another point of view, one can {\it establish a one-to-one correspondence between $\{c_{i}\}$ and $\{d_{i}\}$ fermions}. With this correspondence, the two types of transformations discussed above correspond to {\it an interchange} among the Majorana fermions $\{\eta_{i}\}$ and an {\it orthogonal} transformation on the single-particle Majorana Hamiltonian. On the other hand, a U(1) phase transformation of the complex fermion $c_{i}$ and $d_{i}$ will not change physical properties of the system, provided that the single-particle Hamiltonian changes accordingly. It corresponds to a O(2) rotation of the Majorana fermions. Under the definition $c_{i}^{\dagger}=\frac{1}{2}(\eta_{i}^{\alpha}+i\eta_{i}^{\beta})$, we have $c_{i}^{\dagger}\rightarrow e^{i\theta}c_{i}^{\dagger}$ corresponds to
\begin{equation}
\label{O2rotation}
\left(\begin{array}{c}
\eta_{i}^{\alpha}\\\eta_{i}^{\beta}
\end{array}\right)\rightarrow \left(\begin{array}{cc}
\cos \theta&-\sin \theta\\
\sin\theta & \cos\theta 
\end{array}\right)\left(\begin{array}{c}
\eta_{i}^{\alpha}\\\eta_{i}^{\beta}
\end{array}\right).
\end{equation}  
These two types of transformations on the Majorana fermions indicates that the transformation $U$ on the Nambu spinor does not span the whole unitary group but rather span the orthogonal group of the Majorana fermions. To see this, it is important to note that there is a physical constraint on the unitary transformations imposed by the Nambu spinor structure, namely
\begin{equation}
\hat{\sigma}_{x}\Psi=(\Psi^{\dagger})^{T},\qquad \hat{\sigma}_{x}\Phi=(\Phi^{\dagger})^{T},
\end{equation}
in which $\hat{\sigma}^{x}$ is the Pauli matrix acting on the particle-hole space of the Nambu spinor. Since we have $\Phi=U\Psi$, this means that $\hat{\sigma}_{x}U\Psi=U^{*}(\Psi^{\dagger})^{T}$, which in turn implies the following constraint on the unitary transformation matrix
\begin{equation}
\hat{\sigma}_{x}U\hat{\sigma}_{x}=U^{*}.
\end{equation}
Considering these, we see that the full transformation group in the Nambu spinor space of the complex fermion corresponds to the $O(2N)$ rotation group in the Majorana fermion space. Namely, with the definition $\boldsymbol{\eta}^{\alpha}=(\eta_{1}^{\alpha},\cdots,\eta_{N}^{\alpha})^{T}$, and $\boldsymbol{\eta}^{\beta}=(\eta_{1}^{\beta},\cdots,\eta_{N}^{\beta})^{T}$, we have the correspondence between orthogonal transformation of the Majorana fermions
\begin{equation}
\left(\begin{array}{c}
\boldsymbol{\eta}^{\alpha}\\
\boldsymbol{\eta}^{\beta}
\end{array}\right)
\rightarrow R\left(\begin{array}{c}
\boldsymbol{\eta}^{\alpha}\\
\boldsymbol{\eta}^{\beta}
\end{array}\right), \qquad R\in O(2N)
\end{equation}
and unitary transformation of the complex fermions with constraint
\begin{equation}
\Psi\rightarrow U\Psi,\qquad U\in U(2N),\qquad \hat{\sigma}_{x}U\hat{\sigma}_{x}=U^{*}.
\end{equation}
Further mathematical consideration is needed to confirm such a correspondence; nevertheless, we will refer to these transformation as {\it Majorana orthogonal transformations}.

Obviously the Majorana orthogonal transformations preserve the spectrum of the model. It also preserves the symmetries of the model. Specifically if $\mathcal{H}=\Psi^{\dagger}H\Psi$ has a symmetry $\hat{T}$, such as translational symmetry, then for single particle Hamiltonian we have $[H,\hat{T}]=0$. For the transformed Hamiltonian $\mathcal{H}=\Phi^{\dagger}\tilde{H}\Phi$, $\tilde{H}=UHU^{\dagger}$, there is a corresponding symmetry $\hat{T}'=U\hat{T}U^{\dagger}$, such that $[\tilde{H},\hat{T}']=0$. Therefore the Majorana orthogonal transformations define a {\it duality} between complex fermion models, which is referred to as Gaussian duality in Ref. \onlinecite{cobanera2015}.

\subsection{Zero modes in free fermionic systems} \label{subseczeromode}

We now move on to discuss zero modes in free fermionic systems. For a general fermionic hopping model $\mathcal{H}=\Psi^{\dagger}H\Psi=\Psi_{i}^{\dagger}H_{ij}\Psi_{j}$, if an operator 
\begin{equation}
\zeta=\sum_{i}\lambda_{i}\Psi_{i}
\end{equation}
has the property that $[\zeta,\mathcal{H}]=0$, then the operator $\zeta$ is called a zero mode of the model. If $\zeta=\zeta^{\dagger}$ then $\zeta$ is a {\it Majorana zero mode}; otherwise $\zeta$ is a {\it complex zero mode}. The set $\{\lambda_{i}\}$ is called the {\it wavefunction of the zero mode}, which is a generalized wavefunction in the Majorana space. If the amplitude $|\lambda_{i}|^{2}$ in $\{\lambda_{i}\}$ wavefunction is peaked at some point and decay exponatially with the distance from that point, then the zero mode is point-like localized and the {\it position} of the zero mode can thus be defined \cite{chertkov20}. 

From a complex zero mode $\zeta$ we can construct two Majorana operators which are independent from each other, $\frac{1}{2}(\zeta+\zeta^{\dagger})$ and $\frac{i}{2}(\zeta-\zeta^{\dagger})$. If the model has even number of Majorana zero modes, they can also be paired up into several complex zero modes. Fermionic systems with finite sizes always have even number of Majorana zero modes by physical requirement, but these modes are not necessarily close to each other.  The situation is special when the model has one single Majorana zero mode or an odd number of Majorana zero modes around a certain point and other zero modes located far away from them, this is the case which we will focus on for the rest of the paper.

Next we consider the properties of the zero modes when the system undergoes Majorana orthogonal transformation. For single layer pairing (\ref{singlelayernambu}), if the model has a localized zero mode for $\mathcal{H}(d_{i},d_{i}^{\dagger})$, then it has a corresponding localized zero mode for $\mathcal{H}(c_{i},c_{i}^{\dagger})$. We have wave function transformation $\zeta=\sum_{i}\lambda_{i}\Phi_{i}=\sum_{i}\tilde{\lambda}_{j}\Psi_{j}$, in which $\tilde{\lambda}_{j}=\sum_{i}\lambda_{i}U_{ij}$. For the resulting zero mode to be localized, the transformation itself must be local, namely the Majorana fermions can only be interchanged with other Majorana fermions nearby. For double layer pairing (\ref{doublelayernambu}), it can be shown that if the total Hamiltonian $\mathcal{H}_{A+B}$ has a zero mode then {\it it is equivalent to the fact that at least one of the layers has a corresponding zero mode}. In general Majorana orthogonal transformations preserve the existence of the zero modes, namely if the original model has a zero mode, the final model must have a corresponding zero mode. To summarize, the duality between complex {\it free} fermionic models under Majorana orthogonal transformations and the shared properties between them are illustrated by the following diagram.

\begin{tabular}{ccccc}
	\fbox{\begin{minipage}{1.35cm}
			\text{Complex}\\
			\text{model}
			$\{c_{i}\}$
	\end{minipage}}&\begin{minipage}{1.3cm}
		$\xrightarrow{\text{decoupling}}$\\ $\xleftarrow[\text{pairing}\quad]{}$
	\end{minipage}&\fbox{\begin{minipage}{1.5cm}
			\text{Majorana}\\
			\text{model}
			$\{\eta_{i}^{\alpha},\eta_{i}^{\beta}\}$
	\end{minipage}}&\begin{minipage}{1.4cm}
		$\xrightarrow{\text{pairing}\quad}$\\
		$\xleftarrow[\text{decoupling}]{}$
	\end{minipage}&\fbox{\begin{minipage}{1.35cm}
			\text{Complex}\\
			\text{model}
			$\{d_{i}\}$
	\end{minipage}}\\
	&$\rotatebox{150}{$\xleftarrow{\qquad}$}$&$\rotatebox{90}{$\xleftarrow{\quad}$}$&$\rotatebox{35}{$\xleftarrow{\qquad}$}$&\\
	&&\fbox{\begin{minipage}{1.9cm}
			spectrum,\\
			symmetries,\\
			zero mode $\zeta$
	\end{minipage}}&&
\end{tabular}

\subsection{Majorana hopping models reduced from exactly solvable interacting models} \label{subsecexactlysolvable}

Before moving on, we consider another possible origin of the Majorana hopping models in fermionic systems. Inspired by the Kitaev honeycomb model \cite{Kitaev2006}, we are able to construct certain {\it exactly solvable interacting models} that have a lot of conserved Majorana bilinears and can therefore be reduced to Majorana hopping models \cite{chen2018}. To introduce the model, we construct a lattice that has $n$ bonds connecting to a vertex (site); then we put $n+1$ Majorana fermions on each vertex and denote them by $\eta$ and $\gamma^{\alpha}$, $\alpha=1,\cdots,n$. For $n$ being an odd integer, the number of Majorana fermions on each vertex is even and it is possible to define a local Hilbert space of fermions by pairing up the Majorana fermions on every vertex. We label the bonds of the lattice by $1,\cdots,n$ with each type of bond appears once and only once around each vertex. The Hamiltonian of the interacting Majorana model is given by
\begin{equation}
\label{Hforinteractinglatticemodel}
\mathcal{H}=\sum_{\langle ij\rangle_{\alpha}}it^{\alpha}_{ij}(\eta_{i}\eta_{j})+(-J_{ij}^{\alpha})(\eta_{i}\eta_{j})\gamma_{i}^{\alpha}\gamma_{j}^{\alpha},
\end{equation}
in which $\alpha$ takes the values $1,\cdots, n$ depending on the type of the bond $\langle ij\rangle$. The model has the form of a $t-J$ model for Majorana fermions. It is exactly solvable by noting that the link variables $\gamma^{\alpha}_{i}\gamma_{j}^{\alpha}$ commute with other Majorana fermion bilinears in the Hamiltonian and hence commute with the Hamiltonian itself. The $\gamma_{i}^{\alpha}$ Majorana fermions have no dynamics and we can introduce static $Z_{2}$ variables 
\begin{equation}
\sigma_{ij}^{z}=i\gamma_{i}^{\alpha}\gamma_{j}^{\alpha}.
\end{equation}
The model Hamiltonian (\ref{Hforinteractinglatticemodel}) is then transformed into
\begin{equation}
\mathcal{H}=\sum_{\langle ij\rangle_{\alpha}}i\bigg(t_{ij}^{\alpha}+J_{ij}^{\alpha}\sigma_{ij}^{z}\bigg)\eta_{i}\eta_{j}.
\end{equation}
Once the distribution of the $Z_{2}$ variables $\sigma_{ij}^{z}$ is determined, the model is transformed into a Majorana hopping model for $\eta$ Majorana fermions,  $\mathcal{H}=\sum_{\langle ij\rangle_{\alpha}}iA_{ij}^{\alpha}\eta_{i}\eta_{j}$, with $A_{ij}^{\alpha}=t_{ij}^{\alpha}+J_{ij}^{\alpha}\sigma_{ij}^{z}$. Physical eigenstates of this model contain the distribution of the $Z_{2}$ variables and the corresponding fermionic state, they can be written as $|\psi\rangle_{\text{Phys}}=|\{\sigma_{ij}^{z}\}\rangle\otimes|\eta_{\{\sigma\}}\rangle$. Without the proper Gauss law constraints \cite{fradkinbook,fu20181,fu20182,fu20191}, the model cannot be intepreted as $Z_{2}$ lattice gauge theory, the spectrum of the model thus has a huge degeneracy. Despite that, the discussions on the Majorana hopping models from {\it free} complex models can be brought into this type of models. In particular, the results on Majorana zero modes may be brought to this model with some modifications. Detailed study on this is left for the future.

\section{U(1) topological gauge theory of $p_{x}+ip_{y}$ topological superconductor} \label{seccontinuous}

In this section, we apply the Majorana orthogonal transformation to study the $p_{x}+ip_{y}$ topological supercondutor \cite{Read00} of spinless complex fermion. To this end, we obtain a {\it U(1) topological gauge theory of massive Dirac fermion} for the doubled system, using which we explicitly relate the appearance of Majorana zero modes around vortex cores to the {\it parity anormaly} of the massive Dirac fermion. The purpose of this study is threefold. First, the Majorana zero mode that appears in $p_{x}+ip_{y}$ topological superconductor can be seen as a ``prototype" of the Majorana zero modes in various models \cite{Ivanov01,roy10,tewari07,nishida2010,gurarie07}; second, the discussion illustrates the application of Majorana orthogonal transformation to a {\it continuous} model rather than lattice fermionic models; third, our method is independent of previous theoretical approaches \cite{Read00,roy10,gurarie07,nishida2010,tewari07,Chung07,Cheng09,Mizushima10} for the Majorana zero modes and it results in some new understanding of known physics of Majorana zero modes.

\subsection{Majorana orthorgonal transformation for the $p_{x}+ip_{y}$ topological superconductor}

We start with a general superconducting Hamiltonian of spinless fermion $c$ in the real space,
\begin{eqnarray}
\begin{aligned}
\label{H1}
\mathcal{H}_{F}=&\int d^{2}\boldsymbol{x}d^{2}\boldsymbol{x}' \bigg[c^{\dagger}(\boldsymbol{x})\hat{h}(\boldsymbol{x},\boldsymbol{x}')c(\boldsymbol{x}')+\\&\Delta(\boldsymbol{x},\boldsymbol{x}')c^{\dagger}(\boldsymbol{x})c^{\dagger}(\boldsymbol{x}')+\Delta^{*}(\boldsymbol{x},\boldsymbol{x}')c(\boldsymbol{x}')c(\boldsymbol{x})\bigg].
\end{aligned}
\end{eqnarray}
In the Hamiltonian (\ref{H1}), the first term includes the kinetic energy and the chemical potential terms, it satisfies $\hat{h}^{*}(\boldsymbol{x},\boldsymbol{x}')=\hat{h}(\boldsymbol{x}',\boldsymbol{x})$. Furthermore we assume that $\hat{h}$ is real, so it is even under exchange of coordinates $\hat{h}(\boldsymbol{x},\boldsymbol{x}')=\hat{h}(\boldsymbol{x}',\boldsymbol{x})$. On the other hand, following from fermion statistics, the pairing field $\Delta$ is odd under exchange of coordinates, $\Delta(\boldsymbol{x},\boldsymbol{x}')=-\Delta(\boldsymbol{x}',\boldsymbol{x})$.

For the next step, we decouple the fermionic fields $c(\boldsymbol{x})$ into Majorana fields $\eta^{\alpha}(\boldsymbol{x})$ and $\eta^{\beta}(\boldsymbol{x})$,
\begin{equation}
c^{\dagger}(\boldsymbol{x})=\frac{1}{2}(\eta^{\alpha}(\boldsymbol{x})+i\eta^{\beta}(\boldsymbol{x})), \qquad c(\boldsymbol{x})=\frac{1}{2}(\eta^{\alpha}(\boldsymbol{x})-i\eta^{\beta}(\boldsymbol{x})).
\end{equation}
In terms of these Majorana fields, the Hamiltonian (\ref{H1}) can be written as 
\begin{eqnarray}
\label{H2}
\begin{aligned}
\mathcal{H}_{F}=&\frac{1}{4}\int d^{2}\boldsymbol{x}d^{2}\boldsymbol{x}' \times\\&\bigg\{\big(\hat{h}(\boldsymbol{x},\boldsymbol{x}')+2i\operatorname{Im} \Delta(\boldsymbol{x},\boldsymbol{x}')\big)\eta^{\alpha}(\boldsymbol{x})\eta^{\alpha}(\boldsymbol{x}')\\& +\big(\hat{h}(\boldsymbol{x},\boldsymbol{x}')+2\operatorname{Re} \Delta(\boldsymbol{x},\boldsymbol{x}')\big)i\eta^{\beta}(\boldsymbol{x})\eta^{\alpha}(\boldsymbol{x}') \\&+\big(-\hat{h}(\boldsymbol{x},\boldsymbol{x}')+2\operatorname{Re} \Delta(\boldsymbol{x},\boldsymbol{x}')\big)i\eta^{\alpha}(\boldsymbol{x})\eta^{\beta}(\boldsymbol{x}')\\&+\big(\hat{h}(\boldsymbol{x},\boldsymbol{x}')-2i\operatorname{Im} \Delta(\boldsymbol{x},\boldsymbol{x}')\big)\eta^{\beta}(\boldsymbol{x})\eta^{\beta}(\boldsymbol{x}')\bigg\}.
\end{aligned}
\end{eqnarray}

Now we introduce another copy of the same system, with the same Hamiltonian (\ref{H2}), in which the corresponding Majorana fermion fields are $\tilde{\eta}^{\alpha}$ and $\tilde{\eta}^{\beta}$. The new system has no coupling with the original system and the Hamiltonian can be written as $\tilde{\mathcal{H}}_{F}(\tilde{\eta}^{\alpha},\tilde{\eta}^{\beta})$.
According to (\ref{directadditionoftwohamiltonian}), the Hamiltonian of the doubled systems is $\mathcal{H}_{F}(\eta^{\alpha},\eta^{\beta})\oplus\tilde{\mathcal{H}}_{F}(\tilde{\eta}^{\alpha},\tilde{\eta}^{\beta})$, which acts on the total Hilbert space of the two copies of the system. Applying a Majorana orthorgonal transformation, we can pair up the Majorana fields in a different way and define new complex fermion field 
\begin{equation}
\label{ffermion}
f^{\dagger\mu}(\boldsymbol{x})=\frac{1}{2}(\eta^{\mu}(\boldsymbol{x})+i\tilde{\eta}^{\mu}(\boldsymbol{x})), \qquad \mu=\alpha, \beta.
\end{equation}
Using the fact that $\hat{h}$ is real and even under coordinate exchange and $\Delta$ is odd under coordinate exchange we can write the Hamiltonian $\mathcal{H}_{F}\oplus\tilde{\mathcal{H}}_{F}$ in terms of the new fermion field as
\begin{equation}
\label{H7}
\mathcal{H}_{F}\oplus\tilde{\mathcal{H}}_{F}=\int_{\boldsymbol{x},\boldsymbol{x}'}  \left(\begin{array}{cc}
f^{\alpha\dagger}(\boldsymbol{x}) & f^{\beta\dagger}(\boldsymbol{x})
\end{array}\right)\mathbf{H}(\boldsymbol{x},\boldsymbol{x}')\left(\begin{array}{c}
f^{\alpha}(\boldsymbol{x}')\\f^{\beta}(\boldsymbol{x}')
\end{array}\right),
\end{equation}
in which 
\begin{equation}
\mathbf{H}(\boldsymbol{x},\boldsymbol{x}')=\left(\begin{array}{cc}
2i\operatorname{Im}\Delta&i(2\operatorname{Re}\Delta-\hat{h})\\
i(2\operatorname{Re}\Delta+\hat{h})&-2i\operatorname{Im}\Delta
\end{array}\right).
\end{equation}

To simplify the Hamiltonian, we define a constant unitary matrix $\Lambda$,
\begin{equation}
\label{lambdamatrix}
\Lambda=\frac{1}{\sqrt{2}}\left(\begin{array}{cc}
i&i\\1&-1
\end{array}\right),\qquad \Lambda^{\dagger}\Lambda=\hat{I}.
\end{equation}
Using (\ref{lambdamatrix}) a new set of fermion field $\psi^{\alpha}$ and $\psi^{\beta}$ can be introduced by unitary transformaion 
\begin{equation}
\left(\begin{array}{c}
f^{\alpha}(\boldsymbol{x})\\f^{\beta}(\boldsymbol{x})
\end{array}\right)=\Lambda\left(\begin{array}{c}
\psi^{\alpha}(\boldsymbol{x})\\\psi^{\beta}(\boldsymbol{x})
\end{array}\right).
\end{equation}
In terms of the $\psi$ fermions, the Hamiltonian (\ref{H7}) can be written as
\begin{eqnarray}
\begin{aligned}
\label{HH1}
&\mathcal{H}_{F}\oplus\tilde{\mathcal{H}}_{F}=\int d^{2}\boldsymbol{x}d^{2}\boldsymbol{x}'\times \\&\left(\begin{array}{cc}
\psi^{\alpha\dagger}&\psi^{\beta\dagger}
\end{array}\right)_{\boldsymbol{x}}\left(\begin{array}{cc}
-\hat{h}(\boldsymbol{x},\boldsymbol{x}')&-2\Delta^{*}(\boldsymbol{x},\boldsymbol{x}')\\
2\Delta(\boldsymbol{x},\boldsymbol{x}')&\hat{h}(\boldsymbol{x},\boldsymbol{x}')
\end{array}\right)\left(\begin{array}{c}
\psi^{\alpha}\\\psi^{\beta}
\end{array}\right)_{\boldsymbol{x}'}.
\end{aligned}
\end{eqnarray}

From now on we focus on the case of $p_{x}+ip_{y}$ topological superconductor, in which the pair field in real space with no defect is given by
\begin{equation}
\label{deltaconst}
\Delta_{0}(\boldsymbol{x},\boldsymbol{x}')=\rho \delta^{(2)}(\boldsymbol{x}-\boldsymbol{x}')(i\partial_{x'}-\partial_{y'}),
\end{equation}
in which $\rho$ is a constant complex number. Moreover, we assume that the kinetic energy part is local, which means that it also has a factor of $\delta^{(2)}(\boldsymbol{x}-\boldsymbol{x}')$, namely, $\hat{h}(\boldsymbol{x},\boldsymbol{x}')=\tilde{h}(\boldsymbol{x})\delta^{(2)}(\boldsymbol{x}-\boldsymbol{x}')$.
The expression of $\Delta(\boldsymbol{x},\boldsymbol{x}')$ given by (\ref{deltaconst}) works when $\rho$ is a constant over the real space for $p_{x}+ip_{y}$ superconductors. However, when $\rho=\rho_{\boldsymbol{x}}$ is a function of position, the expression (\ref{deltaconst}) leads to $\Delta_{0}(\boldsymbol{x},\boldsymbol{x}')+\Delta_{0}(\boldsymbol{x}',\boldsymbol{x})=-[(i\partial_{x'}-\partial_{y'})\rho_{\boldsymbol{x}'}]\delta^{(2)}(\boldsymbol{x}-\boldsymbol{x}')$. This contradicts the requirement that $\Delta(\boldsymbol{x},\boldsymbol{x}')+\Delta(\boldsymbol{x}',\boldsymbol{x})\equiv 0$, to remedy this, we have to define a new pairing field,
\begin{equation}
\label{deltanew}
\Delta(\boldsymbol{x},\boldsymbol{x}')=\rho_{\boldsymbol{x}}\delta(\boldsymbol{x}-\boldsymbol{x}')(i\partial_{x'}-\partial_{y'})+\frac{1}{2}[(i\partial_{x}-\partial_{y})\rho_{\boldsymbol{x}}]\delta(\boldsymbol{x}-\boldsymbol{x}').
\end{equation}
In order to study the physics of vortices in the $p_{x}+ip_{y}$ superconductor, we assume that the pairing field $\rho_{\boldsymbol{x}}$ has a constant modulus and a phase which is a function of position. In this situation a {\it gauge field} can be introduced from the phase \cite{hasson20041}, namely
\begin{equation}
\label{definitionofgaugefield}
\rho_{\boldsymbol{x}}=|\rho|e^{i\phi_{\boldsymbol{x}}},\qquad a_{\mu}=\frac{1}{2}\partial_{\mu}\phi_{\boldsymbol{x}}.
\end{equation}
The physical meaning of the gauge field will become clear shortly. The phase $\phi_{\boldsymbol{x}}$ can be written in terms of the gauge field as $\phi_{\boldsymbol{x}}=2\int^{\boldsymbol{x}}a_{\mu}dx^{\mu}$, and the definition leads to $\partial_{\mu}\rho_{\boldsymbol{x}}=2ia_{\mu}\rho_{\boldsymbol{x}}$ and $\partial_{\mu}\rho_{\boldsymbol{x}}^{*}=-2ia_{\mu}\rho^{*}_{\boldsymbol{x}}$. Using the gauge field $a_{\mu}$, we define the {\it covariant derivative} $D_{\mu}=\partial_{\mu}+ia_{\mu}$. With these setup, the Hamiltonian (\ref{HH1}) for the {\it doubled} $p_{x}+ip_{y}$ superconductor can be written as
\begin{equation}
\label{HH3}
\mathcal{H}_{F}\oplus\tilde{\mathcal{H}}_{F}=\int d^{2}\boldsymbol{x} \left(\begin{array}{cc}
\psi^{\alpha\dagger} & \psi^{\beta\dagger}
\end{array}\right)_{\boldsymbol{x}}\mathbf{H}_{p+ip}(\boldsymbol{x})\left(\begin{array}{c}
\psi^{\alpha}\\\psi^{\beta}
\end{array}\right)_{\boldsymbol{x}},
\end{equation}
in which
\begin{equation}
\mathbf{H}_{p+ip}(\boldsymbol{x})=\left(\begin{array}{cc}
-\tilde{h}(\boldsymbol{x})&2\rho^{*}_{\boldsymbol{x}}\big(iD_{x}^{*}+D_{y}^{*}\big)\\
2\rho_{\boldsymbol{x}}\big(iD_{x}-D_{y}\big)&\tilde{h}(\boldsymbol{x})
\end{array}\right).
\end{equation}
In what follows we assume that the $\tilde{h}(\boldsymbol{x})$ terms contain no spatial derivative, which leads to a {\it gauge invariance} for the Hamiltonian (\ref{HH3}). The gauge transformation is given by the following,
\begin{eqnarray}
\begin{aligned}
&\psi^{\alpha}(\boldsymbol{x})\rightarrow e^{i\theta_{\boldsymbol{x}}}\psi^{\alpha}(\boldsymbol{x}),\qquad \psi^{\beta}(\boldsymbol{x})\rightarrow e^{-i\theta_{\boldsymbol{x}}}\psi^{\beta}(\boldsymbol{x});\\ &\rho_{\boldsymbol{x}}\rightarrow e^{-2i\theta_{\boldsymbol{x}}}\rho_{\boldsymbol{x}},\qquad a_{\mu}\rightarrow a_{\mu}-\partial_{\mu}\theta_{\boldsymbol{x}}.
\end{aligned}
\end{eqnarray}
From this one can read off the charges of the three matter field $\psi^{\alpha}$, $\psi^{\beta}$ and $\rho$ as $+1$, $-1$ and $-2$ respectively.

\subsection{Massive Dirac fermion and U(1) topological gauge theory}

To proceed, we use the charge $-1$ $\psi^{\beta}$ and charge $+2$ $\rho^{*}$ to form a charge $+1$ object, after that the Hamiltonian (\ref{HH3}) can be written as a Dirac fermion formulism, provided that the $\tilde{h}(\boldsymbol{x})$ term doesn't contain any spatial derivative. To this end the following equations will be useful
\begin{eqnarray}
\begin{aligned}
&\rho^{*}_{\boldsymbol{x}}D_{\mu}^{*}\psi^{\beta}(\boldsymbol{x})=D_{\mu}\big(\rho_{\boldsymbol{x}}^{*}\psi^{\beta}(\boldsymbol{x})\big), \\ &\rho^{*}_{\boldsymbol{x}}\bigg(iD_{x}^{*}+D_{y}^{*}\bigg)\psi^{\beta}(\boldsymbol{x})=\bigg(iD_{x}+D_{y}\bigg)\big(\rho_{\boldsymbol{x}}^{*}\psi^{\beta}(\boldsymbol{x})\big).
\end{aligned}
\end{eqnarray}
With these, we are able to bring the Hamiltonian (\ref{HH3}) into the following suggestive form
\begin{eqnarray}
\begin{aligned}
\label{HH5}
&\mathcal{H}_{F}\oplus\tilde{\mathcal{H}}_{F}=\\&\int d^{2}\boldsymbol{x} \left(\begin{array}{cc}
\psi^{\alpha\dagger} & e^{i\phi_{\boldsymbol{x}}}\psi^{\beta\dagger}
\end{array}\right)_{\boldsymbol{x}}\tilde{\mathbf{H}}_{p+ip}(\boldsymbol{x})\left(\begin{array}{c}
\psi^{\alpha}\\e^{-i\phi_{\boldsymbol{x}}}\psi^{\beta}
\end{array}\right)_{\boldsymbol{x}},
\end{aligned}
\end{eqnarray}
in which
\begin{equation}
\tilde{\mathbf{H}}_{p+ip}(\boldsymbol{x})=\left(\begin{array}{cc}
-\tilde{h}(\boldsymbol{x})&2|\rho|\big(iD_{x}+D_{y}\big)\\
2|\rho|\big(iD_{x}-D_{y}\big)&\tilde{h}(\boldsymbol{x})
\end{array}\right).
\end{equation}
Now we define another fermion field 
\begin{equation}
\chi^{\beta}(\boldsymbol{x})=e^{-i\phi_{\boldsymbol{x}}}\psi^{\beta}(\boldsymbol{x})=e^{-2i\int^{\boldsymbol{x}}a_{\mu}dx^{\mu}}\psi^{\beta}(\boldsymbol{x}).
\end{equation}
It has the physical intepretation of being the fermion field $\psi^{\beta}$ attached to a half-infinite Wilson line of the gauge field and it has charge $+1$. Then the two fermions $\psi^{\alpha}$ and $\chi^{\beta}$ with the same charge can be paired up into {\it Dirac fermion}
\begin{equation}
\Psi(\boldsymbol{x})=\left(\begin{array}{c}
\psi^{\alpha}(\boldsymbol{x})\\\chi^{\beta}(\boldsymbol{x})
\end{array}\right).
\end{equation}
The Hamiltonian (\ref{HH5}) can thus be written as the Dirac Hamiltonian 
\begin{eqnarray}
\begin{aligned}
\label{HH7}
\mathcal{H}_{F}\oplus\tilde{\mathcal{H}}_{F}=&2|\rho|\int d^{2}\boldsymbol{x}\bigg[\Psi^{\dagger}(\boldsymbol{x})\bigg(-\frac{\tilde{h}(\boldsymbol{x})}{2|\rho|}\bigg)\sigma^{z}\Psi(\boldsymbol{x})+\\&\Psi^{\dagger}(\boldsymbol{x})\bigg(i\sigma^{x}D_{x}+i\sigma^{y}D_{y}\bigg)\Psi(\boldsymbol{x})\bigg],
\end{aligned}
\end{eqnarray}
in which $\sigma^{x}$, $\sigma^{y}$ and $\sigma^{z}$ are Pauli matrices. We choose the following $\gamma$ matrices \cite{son20151}, 
\begin{equation}
\gamma^{0}=\sigma^{z},\qquad \gamma^{1}=i\sigma^{y},\qquad \gamma^{2}=-i\sigma^{x}.
\end{equation}
They satisfy the Clifford algebra $\{\gamma^{\mu},\gamma^{\nu}\}=2\eta^{\mu\nu}$, in which $\eta^{\mu\nu}$ is the metric tensor. Furthermore the conjugate spinor can be defined as $\bar{\Psi}=\Psi^{\dagger}\sigma^{z}=\Psi^{\dagger}\gamma^{0}$. Using these the Hamiltonian (\ref{HH7}) can be written as
\begin{eqnarray}
\begin{aligned}
\label{HH8}
\mathcal{H}_{F}\oplus\tilde{\mathcal{H}}_{F}=&2|\rho|\int d^{2}\boldsymbol{x}\bigg[\bigg(-\frac{\tilde{h}(\boldsymbol{x})}{2|\rho|}\bigg)\bar{\Psi}(\boldsymbol{x})\Psi(\boldsymbol{x})+\\&\bar{\Psi}(\boldsymbol{x})i\bigg(\gamma^{1}D_{x}+\gamma^{2}D_{y}\bigg)\Psi(\boldsymbol{x})\bigg].
\end{aligned}
\end{eqnarray}
The Hamiltonian (\ref{HH8}) describes a {\it massive Dirac fermion}. In the low energy limit we neglect the kinetic energy part in $\tilde{h}(\boldsymbol{x})$ and leave just the chemical potential term, namely $\tilde{h}(\boldsymbol{x})\rightarrow -\mu$, then the Dirac fermion mass is given by
\begin{equation}
\label{massgapdirac}
m_{\Psi}=\frac{\tilde{h}(\boldsymbol{x})}{2|\rho|}\rightarrow -\frac{\mu}{2|\rho|}.
\end{equation}
For $\mu>0$ we have $m_{\Psi}<0$.

For massive Dirac fermion in $(2+1)D$ as in our case, one key result is that the low energy effective theory contains a Chern-Simons term of the gauge field, the appearance of which is referred to as the {\it parity anormaly} \cite{redlich1984,niemi1983,semenoff1984,fradkin1986,jackiw1984,qi2013}. In particular, if the energy scale we are interested in is much smaller than the mass gap (\ref{massgapdirac}), integrating out the Dirac fermion will result in an effective action
\begin{equation}
S_{\text{eff}}=\frac{1}{4\pi}\frac{\operatorname{sgn}(m_{\Psi})}{2}\int d^{3}\boldsymbol{x}\epsilon^{\mu\nu\lambda}a_{\mu}\partial_{\nu}a_{\lambda}.
\end{equation}
This action has level $k=\pm\frac{1}{2}$ and thus is not gauge invariant \cite{fradkinbook,freedman04}. To solve this problem we have to perform proper regularization of the theory, after which the effective action becomes
\begin{equation}
\label{chernsimons}
S_{\text{eff}}=\frac{1}{4\pi}\frac{1}{2}\big[\operatorname{sgn}(m_{\Psi})-1\big]\int d^{3}\boldsymbol{x}\epsilon^{\mu\nu\lambda}a_{\mu}\partial_{\nu}a_{\lambda}.
\end{equation}
Nonzero action (\ref{chernsimons}) requires that $m_{\Psi}$ being negative, this is equivalent to chemical potential $\mu>0$, namely the condition of the {\it strong pairing phase} \cite{Read00}. In the strong pairing phase, the resulting level of the Chern-Simons action is $k=-1$. For the doubled system, a full vortex corresponds to a $\pi$ flux of the gauge field. The Chern-Simons term poses a constraint on the charge and flux in the system; consequencely, the $\pi$ flux will have a charge $\pm \frac{1}{2}$ attached to it, this {\it degeneracy} of positive and negative charge corresponds to {\it one and only one complex zero mode} for the $f$ fermion of the doubled system. In particular, suppose that the complex zero mode is $\zeta(\gamma_{\boldsymbol{r}},\tilde{\gamma}_{\boldsymbol{r}})$, which can be written as a linear function of $\eta^{\alpha,\beta}$ and $\tilde{\eta}^{\alpha,\beta}$, and $\gamma_{\boldsymbol{r}}(\eta)$ and $\tilde{\gamma}_{\boldsymbol{r}}(\tilde{\eta})$ are the component of the complex zero mode from each layer. One can see that $\gamma_{\boldsymbol{r}}$ and $\tilde{\gamma}_{\boldsymbol{r}}$ must be Majorana, otherwise if one of them was complex then there would be at least two complex zero modes. In light of this $\gamma_{\boldsymbol{r}}$ is the Majorana zero mode associated with the vortex in a single-layer $p+ip$ superconductor. Therefore we conclude that each $p_{x}+ip_{y}$ superconductor layer should have a {\it Majorana zero mode} around the corresponding vortex.  

\subsection{Application: splitting of degeneracy of multiple vortices in the $p_{x}+ip_{y}$ topological superconductor}

We have built a two-dimensional U(1) gauge theory of massive Dirac fermion for doubled $p_{x}+ip_{y}$ superconductors. When there are multiple vortices in the single-layer superconductor, the corresponding doubled system will have multiple fluxes, each come with a charge on its core. There is electromagnetic interaction between these charges under the U(1) gauge theory. Such interaction will lead to splitting of degeneracy for the multiple vortices configuration of the $p_{x}+ip_{y}$ superconductor.

To determine the magnitude of such splitting, we first have to complete the gauge field dynamics part of the U(1) gauge theory. From the general theory of superconductor, there is a non-linear sigma model term in the Hamiltonian for phase fluctuation
\begin{equation}
\mathcal{H}_{s}=\frac{r_{s}}{2}\big(\nabla\phi_{\boldsymbol{x}}\big)^{2},
\end{equation}
in which the $r_{s}$ is the superconductor stiffness. In terms of the gauge field (\ref{definitionofgaugefield}) this term can be written as
\begin{equation}
\label{masstermofgaugefield}
\mathcal{H}_{s}=2r_{s}a_{\mu}^{2},
\end{equation}
which becomes a mass term for the U(1) gauge field. The U(1) gauge theory is thus in a {\it Higgs phase} and the gauge symmetry is broken to $Z_{2}$. Because we have two identical layers, the mass term of the gauge field should be double that of (\ref{masstermofgaugefield}).

To proceed and compute the Columb interaction between charges, we approximately treat $a_{\mu}$ as a scalar field $b$. The Lagrangian for the scalar field is given as
\begin{equation}
\mathcal{L}=\frac{1}{2}\partial_{\mu}b\partial^{\mu}b-\frac{1}{2}m_{b}^{2}b^{2},
\end{equation} 
the mass of the scalar field is the same as the gauge field $m_{b}=2\sqrt{2r_{s}}$. A kinetic energy term is added to account for the (possible) {\it Maxwell term} of the original gauge field $a_{\mu}$. We then make the following approximation for the fermion interaction vertex
\begin{equation}
\bar{\psi}\gamma^{\mu}a_{\mu}\psi\rightarrow b\bar{\psi}\psi.
\end{equation}
The interaction is thus described by a Yukawa potential in 2D. By Fourier transformation one can write down the potential in real space, with distance between charges given by $R$,
\begin{equation}
V(R)\sim e^{2}\int \frac{d^{2}\boldsymbol{p}}{(2\pi)^{2}}\frac{e^{i\boldsymbol{p}\cdot\boldsymbol{R}}}{p^{2}+m_{b}^{2}}=\frac{e^{2}}{2\pi}K_{0}(m_{b}R),
\end{equation}
in which $K_{0}$ denotes the {\it modified Bessel function} and $e$ is the unit charge. It is noteworthy that the asymptotic behaviour of the modified Bessel function is  $K_{0}(x)\rightarrow \sqrt{\frac{\pi}{2}}\frac{e^{-x}}{\sqrt{x}}$ as $x\rightarrow \infty$.

Now we consider two complex zero modes $\zeta(\gamma_{\boldsymbol{r}},\tilde{\gamma}_{\boldsymbol{r}'})$ and $\zeta'(\gamma_{\boldsymbol{r}'},\tilde{\gamma}_{\boldsymbol{r}}')$.  Their charges under the U(1) gauge field and the interaction between them is determined by the filling of these two modes. Since the charges of these modes are $\pm\frac{1}{2}$, the interaction is given by $\frac{1}{4}V(R)$ computed above if both of them are filled or both are empty. Otherwise if one is filled the other empty, the interaction is $-\frac{1}{4}V(R)$. In other words, the interaction is determined by the {\it fermion parity} of the two complex fermion, which is propotional to the product of Majorana fermions $\gamma_{\boldsymbol{r}}\tilde{\gamma}_{\boldsymbol{r}}\gamma_{\boldsymbol{r}'}\tilde{\gamma}_{\boldsymbol{r}'}$. If the four Majorana zero modes are paired within each layer, the formula of the fermion parity remains the same. When we keep the filling condition of the complex fermion on the second layer and change the filling condition of the first layer, the fermion parity of the doubled system changes sign and the interaction energy changes by $\frac{1}{2}V(R)$. This leads to the conclusion that the degeneracy of the two Majorana zero modes within a single-layer $p_{x}+ip_{y}$ superconductor is splited by $\frac{1}{2}V(R)$ because of the interaction effect. This result obtained from field theory perspective qualitatively agrees with previous results \cite{Chung07,Cheng09,Mizushima10} which adopted a wave-function approach.

Some discussion is in order about the topological superconductor. First, in the doubled $p_{x}+ip_{y}$ superconductor system, a one dimensional domain wall between $\mu>0$ and $\mu<0$ phases will have a {\it complex} 1D mode associated with it. For single layer system, the corresponding 1D mode is {\it Majorana}, this can be intepreted as the {\it edge state} of the $p_{x}+ip_{y}$ superconductor. Second, a one dimensional $p$-wave superconductor can be obtained from the 2D $p_{x}+ip_{y}$ topological superconductor by {\it dimensional reduction} \cite{fradkinbook,qi11,qi2008}. Specifically on $y$-direction, the system is folded into a cylinder with radius $\tilde{r}\rightarrow 0$. The fermionic field $c(\boldsymbol{x})$ is replaced by $c_{n}(x)$, which are the Fourier modes on $y$-direction. As the radius $\tilde{r}\rightarrow 0$ the only mode left in the spectrum is $n=0$. The effective Chern-Simons action for doubled system (\ref{chernsimons}) becomes \cite{fradkinbook}
\begin{equation}
S_{\text{eff}}=\frac{1}{2\pi}\frac{1}{2}\big[\operatorname{sgn}(m_{\Psi})-1\big]\int dxdt \Phi(x,t)\epsilon^{\mu\nu}\partial_{\mu}a_{\nu},
\end{equation}
in which $\Phi(x,t)=\oint dy a_{2}$ is the flux. In the 1D $p$-wave superconductor, there are two types of defects which can have fermionic mode bound to it; one is the $\mu$ defect which corresponds to the domain wall of the 2D system mentioned before; the second one is the $\Delta$ defect created by a gradiant of $\Phi$. We will discuss these in detail within a lattice model of 1D $p$-wave superconductor in the following sections. Looking forward, a doubled $(p_{x}+ip_{y})\times(p_{x}-ip_{y})$ topological superconductor system \cite{qi2009} should have a parity and time-reversal invariant $U_{1}(1)\times\bar{U}_{1}(1)$ Chern-Simons effective field theory \cite{freedman04}, details of which is left for future study.

\section{Zero mode in lattice Majorana hopping models: building blocks} \label{secsimplemodels}

We now turn to discuss Majorana zero modes in 1D and 2D {\it lattice} Majorana hopping models. We focus mostly on infinite lattices. For a general Majorana hopping model, whose Hamiltonian is given by
\begin{equation}
\mathcal{H}=\sum_{ij}\frac{1}{2}A_{ij}i\eta_{i}\eta_{j},\qquad A_{ij}=-A_{ji}.
\end{equation}
If there is a zero mode $\zeta=\sum_{i}\lambda_{i}\eta_{i}$, with real-space wave function $\{\lambda_{i}\}$, then we have $\bigg[\sum_{i}\lambda_{i}\eta_{i},\sum_{jk}\frac{1}{2}A_{jk}i\eta_{j}\eta_{k}\bigg]=0$; the solution to this equation is obtained by requiring that for each of the Majorana fermion the coefficient is zero, which leads to $\sum_{i}A_{ki}\lambda_{i}=0$. This is an eigenvector equation for matrix $\{A_{ij}\}$ with eigenvalue zero. Since we are focusing on zero modes bound to point-like defects, an important requirement is the {\it normalization condition}. For wavefunction $\{\lambda_{i}\}$, we require that $\sum_{i}|\lambda_{i}|^{2}$ converges to a finite value, only such solution represents a defect zero mode.

For all the Majorana hopping models, the Majorana fermions and their sites can always be divided into two groups $\alpha$ and $\beta$, the Majorana fermions are labelled by $\eta_{i_{\alpha}}$ and $\eta_{i_{\beta}}$. Correspondingly the hopping coefficients are also divided into $A_{i_{\alpha}j_{\alpha}}$, $A_{i_{\beta}j_{\beta}}$ and $A_{i_{\alpha}j_{\beta}}$, so that the Hamiltonian is given by
\begin{eqnarray}
\begin{aligned}
\mathcal{H}=&\sum_{i_{\alpha}j_{\beta}}A_{i_{\alpha}j_{\beta}}i\eta_{i_{\alpha}}\eta_{j_{\beta}}+\frac{1}{2}\sum_{i_{\alpha}j_{\alpha}}A_{i_{\alpha}j_{\alpha}}i\eta_{i_{\alpha}}\eta_{j_{\alpha}}\\&+\frac{1}{2}\sum_{i_{\beta}j_{\beta}}A_{i_{\beta}j_{\beta}}i\eta_{i_{\beta}}\eta_{j_{\beta}}.
\end{aligned}
\end{eqnarray}
For some Majorana hopping models, there is a {\it possible dividing} such that all $A_{i_{\alpha}j_{\alpha}}\equiv 0$ and $A_{i_{\beta}j_{\beta}}\equiv 0$, and the Hamiltonian becomes 
\begin{equation}
\label{twogroupshamiltonian}
\mathcal{H}=\sum_{i_{\alpha}j_{\beta}}A_{i_{\alpha}j_{\beta}}i\eta_{i_{\alpha}}\eta_{j_{\beta}}.
\end{equation}
In this situation, we can have separate zero modes $\zeta^{\alpha}=\sum_{i_{\alpha}}\lambda_{i_{\alpha}}\eta_{i_{\alpha}}$ and $\zeta^{\beta}=\sum_{i_{\beta}}\lambda_{i_{\beta}}\eta_{i_{\beta}}$.
The equation for the wave function of the zero modes become
\begin{equation}
\label{twogroupszeromodes}
\sum_{i_{\alpha}}A_{i_{\alpha}k_{\beta}}\lambda_{i_{\alpha}}=0,\qquad \sum_{i_{\beta}}A_{k_{\alpha}i_{\beta}}\lambda_{i_{\beta}}=0.
\end{equation}
This type of Majorana hopping models are referred to as {\it simple models}. Some local Majorana orthogonal transformations can bring the simple models into slightly more complex form.        Simple models and these generalizations form the {\it building blocks} of more complex models with Majorana zero modes. In the following, we discuss some of these building blocks in both one and two dimensions.

\subsection{Zero mode in 1D simple models}

For 1D lattices, the wavefunction of the zero mode $\{\lambda_{i}\}$ forms a {\it real number sequence}. For simple models, the equation for the wavefunction (\ref{twogroupszeromodes}) determine the {\it recurrence relation} of the number sequence. Each recurrence relation is represented by a {\it characteristic equation}, the order of which can be used to classified the zero mode. Here we discuss two models, the first order model from the decoupled Su-Schrieffer-Heeger (SSH) model \cite{su791,su801} and the second order Kitaev chain \cite{Kitaev2001}. 

\subsubsection{First order zero mode} \label{subsecssh}

\begin{figure}
\includegraphics[width=0.4\textwidth]{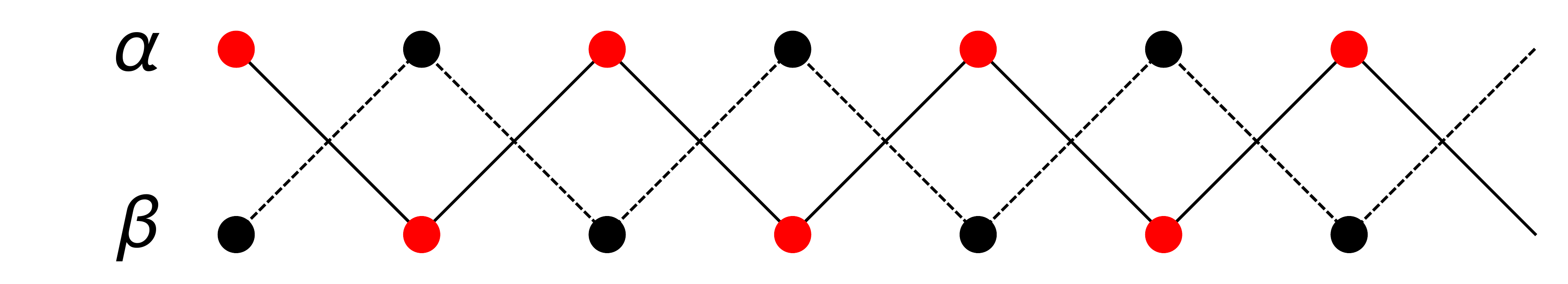}
\caption{The Majorana representation of the Su-Schrieffer-Heeger (SSH) model. The upper dots represent Majorana fermion $\eta_{i}^{\alpha}$ and the lower dots represent Majorana fermion $\eta_{i}^{\beta}$. The model decouples into two independent Majorana hopping models denoted by red dots with solid lines and black dots with dashed lines.}
\label{figssh1}
\end{figure}

\begin{figure}
\includegraphics[width=0.4\textwidth]{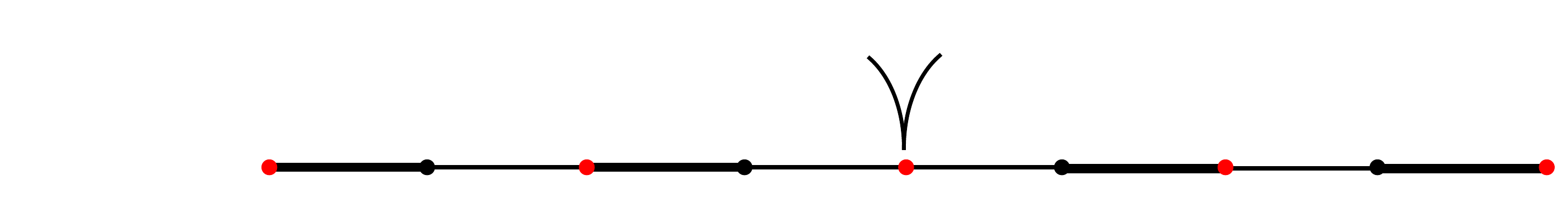}
\caption{The defect Majorana zero mode in one of the {\it decoupled} Su-Schrieffer-Heeger (SSH) model. The magnitudes of the Majorana hopping coefficients are denoted by thick and thin lines respectively. The location of the zero mode is marked by the arrow. The nonzero components of the wavefunction $\lambda_{i}$ are denoted by red dots.}
\label{figssh2}
\end{figure}

To discuss the zero modes, we begin by considering the simpliest 1D complex fermion hopping model, whose Hamiltonian is given by Eq. (\ref{generalcomplexhamiltonian}) with all the hopping coefficients $t_{i,j}$ assumed to be real and all the pairing $\Delta_{ij}=0$. After decoupling the complex fermions into Majorana fermions, the Hamiltonian can be written as
\begin{equation}
\mathcal{H}_{c}=\sum_{i}t_{i,i+1}\frac{i}{2}(\eta_{i}^{\beta}\eta_{i+1}^{\alpha}-\eta_{i}^{\alpha}\eta_{i+1}^{\beta}).
\end{equation}
The model automatically decouples into two separate Majorana hopping chains which do not talk to each other, as shown by Fig. \ref{figssh1}. In each of the two Majorana hopping models, the Majorana breaks into two groups and the Hamiltonian satisfies the ``simple model" condition as illustrated by (\ref{twogroupshamiltonian}). Now we take one of the chains 
\begin{equation}
\label{sshmodelmajoranah}
\mathcal{H}_{\eta}=-\frac{1}{2}\sum_{k}\bigg(t_{2k,2k+1}i\eta_{2k}^{\alpha}\eta_{2k+1}^{\beta}-t_{2k+1,2k+2}i\eta_{2k+1}^{\beta}\eta_{2k+2}^{\alpha}\bigg).
\end{equation}
From the equations for zero modes (\ref{twogroupszeromodes}), we have the wave function of the zero mode $\zeta^{\alpha}$ satisfies {\it first order recurrence relation} $t_{2k,2k+1}\lambda^{\alpha}_{2k}+t_{2k+1,2k+2}\lambda^{\alpha}_{2k+2}=0$, which can be solved by
\begin{equation}
\frac{\lambda^{\alpha}_{2k+2}}{\lambda^{\alpha}_{2k}}=-\frac{t_{2k,2k+1}}{t_{2k+1,2k+2}}.
\end{equation}
As long as 
\begin{eqnarray}
\begin{aligned}
\label{sshcondition}
&\bigg\arrowvert\frac{t_{2k,2k+1}}{t_{2k+1,2k+2}}\bigg\arrowvert\leq \delta<1,\qquad \text{for all } k>0; \\ &\bigg\arrowvert\frac{t_{2k-1,2k}}{t_{2k-2,2k-1}}\bigg\arrowvert\leq \delta<1,\qquad \text{for all } k<0,
\end{aligned}
\end{eqnarray}
the wavefunction $\{\lambda_{i}\}$ is normalizable and peaked at position $k=0$, it decays exponatially with the distance to $k=0$ (see Fig. \ref{figssh2}). There is another zero energy solution for $\eta^{\beta} :\{\lambda_{i}^{\beta}\}$ which is {\it not normalizable}. So the model (\ref{sshmodelmajoranah}) under condition (\ref{sshcondition}) has a single Majorana zero mode at $k=0$.

The complex fermion model under condition (\ref{sshcondition}) is actually a Su-Schrieffer-Heeger (SSH) model with defect \cite{su791,su801}. The model is decoupled into two independent Majorana layers, each of which carries a Majorana zero mode around the defect. The complex fermion SSH model thus has a {\it complex zero mode}.

\subsubsection{Second order zero mode in the Kitaev chain} \label{subsectionkitaevchain}

\begin{figure}
\includegraphics[width=0.4\textwidth]{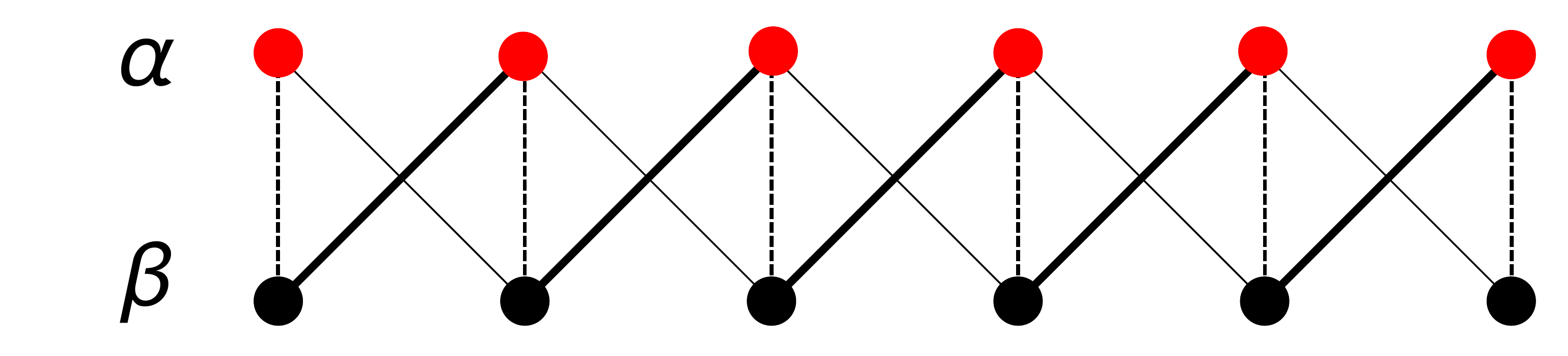}
\caption{The Majorana representation of the Kitaev chain. The upper dots represent Majorana fermion $\eta_{i}^{\alpha}$ and the lower dots represent Majorana fermion $\eta_{i}^{\beta}$. The different magnitudes of the Majorana hopping coefficients on different bonds are denoted by thin solid lines, thick solid lines and dashed lines respectively.}
\label{figkitaev}
\end{figure}

We now turn to discuss a Majorana hopping model whose zero mode recurrence relation has a second order characteristic equation, namely the Kitaev chain \cite{Kitaev2001}. There are a lot of studies focusing on the edge modes of the finite or semi-finite Kitaev chain, here in this section, we consider the {\it defect zero modes} in an {\it infinite Kitaev chain}. The Kitaev chain has the Hamiltonian as follows
\begin{equation}
\label{kitaevchainhamiltonian}
\mathcal{H}=\sum_{i}t_{i}(c_{i}^{\dagger}c_{i+1}+c_{i+1}^{\dagger}c_{i})+\Delta_{i}(c_{i}^{\dagger}c_{i+1}^{\dagger}+c_{i+1}c_{i})+\mu_{i}(c_{i}^{\dagger}c_{i}-\frac{1}{2}).
\end{equation}
In the Hamiltonian, $t_{i}$ and $\Delta_{i}$ are assumed to be real and $t_{i}>0$. To study the model, we start by decoupling the complex fermion into two Majorana fermions, $c_{i}^{\dagger}=\frac{1}{2}(\eta_{i}^{\alpha}+i\eta_{i}^{\beta})$. The Hamiltonian (\ref{kitaevchainhamiltonian}) can thus be written in terms of the Majorana fermions
\begin{equation}
\label{kitaevchainmajoranah}
\mathcal{H}=\sum_{i}\frac{1}{2}(t_{i}+\Delta_{i})i\eta_{i}^{\beta}\eta_{i+1}^{\alpha}+\frac{1}{2}(\Delta_{i}-t_{i})i\eta_{i}^{\alpha}\eta_{i+1}^{\beta}+\frac{1}{2}\mu_{i}i\eta_{i}^{\beta}\eta_{i}^{\alpha}.
\end{equation}
Unlike the SSH model, the Majorana Hamiltonian does not decouple into two independent layers. As shown by Fig. \ref{figkitaev}, the Hamiltonian (\ref{kitaevchainmajoranah}) satisfies the ``simple model" condition, hence we can have separate possible zero modes for $\eta^{\alpha}$ and $\eta^{\beta}$. Specifically for zero mode $\zeta^{\alpha}=\sum_{k}\lambda^{\alpha}_{k}\eta_{k}^{\alpha}$, we have the following recurrence relation based on (\ref{twogroupszeromodes})
\begin{equation}
\label{zetaalpha}
(t_{k}+\Delta_{k})\lambda_{k+1}^{\alpha}+\mu_{k}\lambda_{k}^{\alpha}-(\Delta_{k-1}-t_{k-1})\lambda_{k-1}^{\alpha}=0.
\end{equation}
Similarly for the other zero mode $\zeta^{\beta}=\sum_{k}\lambda_{k}^{\beta}\eta_{k}^{\beta}$, the recurrence relation for the wavefunction is given by
\begin{equation}
\label{zetabeta}
(\Delta_{k}-t_{k})\lambda_{k+1}^{\beta}-\mu_{k}\lambda_{k}^{\beta}-(\Delta_{k-1}+t_{k-1})\lambda_{k-1}^{\beta}=0.
\end{equation}

For the moment, we assume that the $t_{k}$ and $\Delta_{k}$ and $\mu_{k}$ are constants that do not depend on the position $k$. The recurrence relation (\ref{zetaalpha}) can be brought into the form
\begin{equation}
\label{recurrenceforalpha}
\lambda_{k+1}^{\alpha}-x_{1}^{\alpha}\lambda_{k}^{\alpha}=x_{2}^{\alpha}(\lambda_{k}^{\alpha}-x_{1}^{\alpha}\lambda_{k-1}^{\alpha}),
\end{equation}
in which $x_{1,2}^{\alpha}$ are the two solutions of the second order characteristic equation
\begin{equation}
\label{characteristicforalpha}
(\Delta+t)x^{2}+\mu x-(\Delta-t)=0.
\end{equation}
Also the recurrence relation (\ref{zetabeta}) can be brought into the following 
\begin{equation}
\lambda_{k+1}^{\beta}-x_{1}^{\beta}\lambda_{k}^{\beta}=x_{2}^{\beta}(\lambda_{k}^{\beta}-x_{1}^{\beta}\lambda_{k-1}^{\beta}),
\end{equation}
in which $x_{1,2}^{\beta}$ are the two solutions of characteristic equation
\begin{equation}
\label{characteristicforbeta}
(\Delta-t)x^{2}-\mu x-(\Delta+t)=0.
\end{equation}
The solutions for the characteristic equations (\ref{characteristicforalpha}) and (\ref{characteristicforbeta}) are denoted by $x_{\pm}^{\alpha}$ and $x_{\pm}^{\beta}$ respectively, they can be obtained easily and these solutions satisfy the following relations $x_{+}^{\alpha}x_{+}^{\beta}=1$ and $x_{-}^{\alpha}x_{-}^{\beta}=1$.

For the normalization of the zero modes, it is important to determine whether each of these roots $|x_{\pm}^{\alpha,\beta}|$ is greater or smaller than 1. To this end we have the following results, when $|\mu|<2t$ and $\Delta>0$, we have $|x_{\pm}^{\alpha}|<1$ and $|x_{\pm}^{\beta}|>1$; when $|\mu|<2t$ and $\Delta<0$, we have $|x_{\pm}^{\alpha}|>1$ and $|x_{\pm}^{\beta}|<1$. On the other hand, for $|\mu|>2|t|$, both $|x_{\pm}^{\alpha}|$ and $|x_{\pm}^{\beta}|$ have one greater than 1 and the other smaller than 1. Kitaev showed that the existence of boundary modes requires either $|x_{\pm}^{\alpha}|<1$ or $|x_{\pm}^{\beta}|<1$.    Because the model has symmetry under $\alpha\rightarrow \beta$, $t\rightarrow -t$ and $\mu\rightarrow -\mu$, we therefore conclude that if $|\mu|<2|t|$, the model has Majorana boundary mode, such phase is thus referred to as {\it topological}. Conversely, the phase $|\mu|>2|t|$ is non-topological \cite{Kitaev2001}. 

With these results we turn to discuss the defect zero modes in the Kitaev chain created by varying the chemical potential $\mu$. In such defect, the chemical potential $\mu$ takes different values for region $k>0$ and $k<0$; in particular we assume for $k>0$ domain, $|\mu_{+}|<2t$ and the system is in topological phase, and for $k<0$ domain, $|\mu_{-}|>2t$ and the system is in non-topological phase. Both $\mu_{+}$ and $\mu_{-}$ are constants in corresponding domains and we assume that $\Delta>0$ for the entire system. For $k>0$ domain the characteristic equation for $\zeta^{\alpha}$ to the right has two roots $|x_{\pm}^{\alpha}|<1$. From the recurrence equation (\ref{recurrenceforalpha}), assuming $x_{+}^{\alpha}\neq x_{-}^{\alpha}$, the wavefunction sequence can be obtained and the $\lambda_{k}^{\alpha}$ for $k>0$ are fully determined by $\lambda_{1}^{\alpha}$ and $\lambda_{0}^{\alpha}$. For zero mode $\zeta^{\beta}$, since $|x_{\pm}^{\beta}|>1$, the normalization condition cannot be met, so $\zeta^{\beta}$ does not represent a zero mode. For $k<0$ domain, the characteristic equation of $\zeta^{\alpha}$ to the left has one root $|x_{0}^{\alpha}|<1$ and the other root $|\tilde{x}_{0}^{\alpha}|>1$. From (\ref{recurrenceforalpha}) we have 
\begin{eqnarray}
\begin{aligned}
\lambda_{-k-1}^{\alpha}-\tilde{x}_{0}^{\alpha}\lambda_{-k}^{\alpha}=&(x_{0}^{\alpha})^{k}(\lambda_{-1}^{\alpha}-\tilde{x}_{0}^{\alpha}\lambda_{0}^{\alpha}),\\
\lambda_{-k-1}^{\alpha}-x_{0}^{\alpha}\lambda_{-k}^{\alpha}=&(\tilde{x}_{0}^{\alpha})^{k}(\lambda_{-1}^{\alpha}-\tilde{x}_{0}^{\alpha}\lambda_{0}^{\alpha}).
\end{aligned}
\end{eqnarray}
Because $|\tilde{x}_{0}^{\alpha}|>1$, for the mode to be normalizable, one must have $\lambda_{-1}^{\alpha}-x_{0}^{\alpha}\lambda_{0}^{\alpha}=0$, and $\lambda_{-k}^{\alpha}=(x_{0}^{\alpha})^{k}\lambda_{0}^{\alpha}$. The boundary condition at $k=0$ is given by 
\begin{equation}
(t+\Delta)\lambda_{1}^{\alpha}+\mu_{0}\lambda_{0}^{\alpha}+(t-\Delta)\lambda_{-1}^{\alpha}=0,
\end{equation}
in which $\mu_{0}$ is the value of $\mu$ at $k=0$. These determine $\lambda_{\pm 1}$ in terms of $\lambda_{0}$ by 
\begin{equation}
\lambda_{-1}^{\alpha}=x_{0}^{\alpha}\lambda_{0}^{\alpha},\qquad \lambda_{1}^{\alpha}=\frac{(\Delta-t)x_{0}^{\alpha}-\mu_{0}}{t+\Delta}\lambda_{0}^{\alpha}.
\end{equation}
Since this is the only solution, we conclude that {\it there is one Majorana zero mode bound with the $\mu$ defect}.

Before moving on we point out that the first order decoupled SSH model is a special case of the second order Kitaev chain with $\Delta=\pm t$. It is also possible to define higher order models with next-nearest neighbour hopping. 

\subsection{Zero mode in generalized 1D models: the rotated Kitaev chain} \label{subsecrotated}

So far, all the building blocks of the zero mode models that we have considered are simple models. Here we move on to a Majorana hopping model that is not a simple model. In other words, its Majorana fermions cannot be separated into two groups with the hopping paths only connecting Majorana fermion from one group to that of the other group. As discussed in Sec. \ref{SecMOT}, a local U(1) phase rotation of the complex fermions does not change the physical properties of the model. In the Majorana basis, such phase rotation corresponds to local O(2) rotation of the Majorana fermions (\ref{O2rotation}), hence it can alter the Majorana hopping paths and bring a simple model into a complex one. 

\begin{figure}
\includegraphics[width=0.4\textwidth]{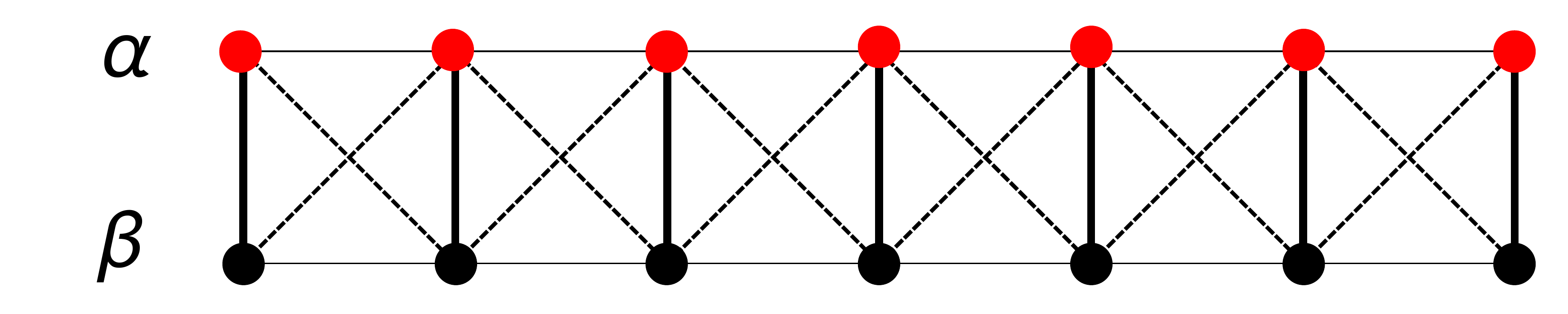}
\caption{The Majorana representation of the {\it rotated Kitaev chain}. The upper dots represent Majorana fermion $\gamma_{i}^{\alpha}$ and the lower dots represent Majorana fermion $\gamma_{i}^{\beta}$. The different magnitudes of the Majorana hopping coefficients on different bonds are denoted by thin solid lines, thick solid lines and dashed lines respectively.}
\label{figrotated}
\end{figure}

In this section, we consider the simplest case with the Kitaev chain. Starting with the Kitaev chain Hamiltonian (\ref{kitaevchainmajoranah}) one can perform a local Majorana orthogonal transformation for all the Majorana fermion, 
\begin{equation}
\left(\begin{array}{c}
\eta_{i}^{\alpha}\\\eta_{i}^{\beta}
\end{array}\right)=\left(\begin{array}{cc}
\cos \theta&\sin\theta\\-\sin\theta&\cos\theta
\end{array}\right)\left(\begin{array}{c}
\gamma_{i}^{\alpha}\\\gamma_{i}^{\beta}
\end{array}\right),
\end{equation}
with $\theta$ being a constant angle and $\gamma_{i}^{\alpha,\beta}$ being a new set of Majorana fermions. If the new Majorana fermions are paired up in the same way $\tilde{c}_{i}^{\dagger}=\frac{1}{2}(\gamma_{i}^{\alpha}+i\gamma_{i}^{\beta})$ the resulting complex Hamiltonian for $\tilde{c}_{i}$ has the same form as the original one (\ref{kitaevchainhamiltonian}) but with $\Delta_{i}$ being a complex number. In terms of the new Majorana fermions $\gamma_{i}^{\alpha}$ and $\gamma_{i}^{\beta}$, the rotated Kitaev chain Hamiltonian can be written as
\begin{eqnarray}
\label{rotatedkitaevchain}
\begin{aligned}
\mathcal{H}=\sum_{i}&\frac{1}{2}(\Delta_{i}\sin 2\theta) i\big(\gamma_{i}^{\beta}\gamma_{i+1}^{\beta}-\gamma_{i}^{\alpha}\gamma_{i+1}^{\alpha}\big)+\\&\frac{1}{2}(t_{i}+\Delta_{i}\cos 2\theta)i\gamma_{i}^{\beta}\gamma_{i+1}^{\alpha}+\\&\frac{1}{2}(\Delta_{i}\cos 2\theta-t_{i})i\gamma_{i}^{\alpha}\gamma_{i+1}^{\beta}+\frac{1}{2}\mu_{i}i\gamma_{i}^{\beta}\gamma_{i}^{\alpha}.
\end{aligned}
\end{eqnarray}
The Hamiltonian can be called {\it the rotated Kitaev chain}. It has the same properties with the original Kitaev chain (\ref{kitaevchainmajoranah}), in particular the topological phase is given by condition $|\mu|<2|t|$ and the Majorana zero mode wave function can be obtained easily from the original one using the transformation. The Majorana zero mode in rotated Kitaev chain involves both $\gamma^{\alpha}$ and $\gamma^{\beta}$ and the model is no longer a simple model, the Majorana hopping path is given in Fig. \ref{figrotated}. 

\subsection{Zero mode in a simple Majorana hopping model on 2D square lattice} \label{subsec2dlattice}

\begin{figure}
\includegraphics[width=0.4\textwidth]{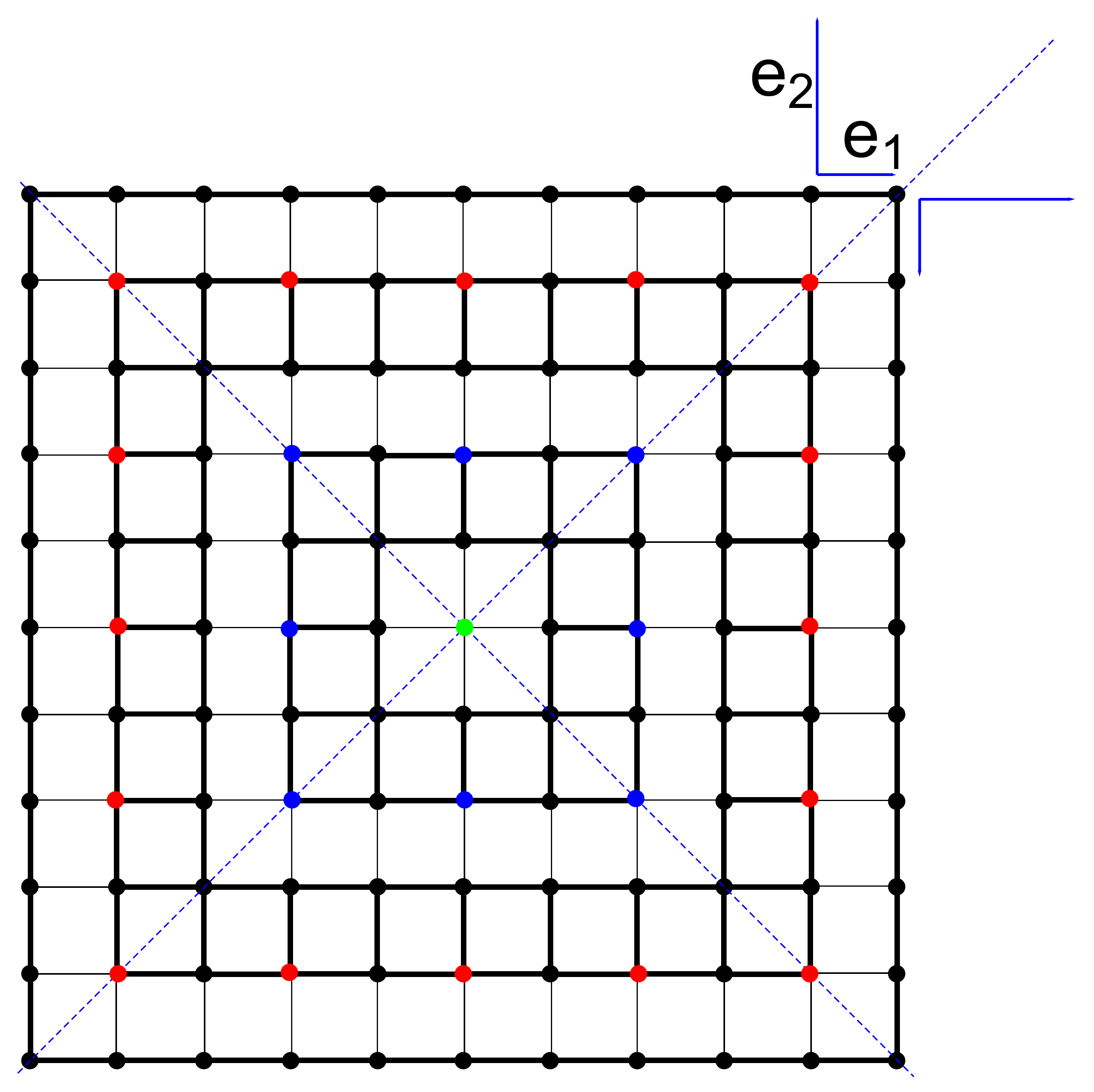}
\caption{The 2D square lattice Majorana hopping model that realizes the vortex of the $p_{x}+ip_{y}$ topological superconductor. The thickness of the lines indicate the magnitude of the Majorana hopping coefficients. The four regions of the lattices are related by rotation of $\frac{\pi}{2}$ and are separated by the blue dashed lines. The lattice vectors $\boldsymbol{e}_{1}$ and $\boldsymbol{e}_{2}$ of two regions are given by the blue arrows. The sites on which the Majorana zero mode wavefunction is nonzero are given by the colored dots. The color of the site indicates the magnitude of the wavefunction on this site, whose ratios are determined by the magnitude of the hopping coefficients.}
\label{fig2d}
\end{figure}

Now we move on to discuss lattice Majorana hopping models that can host zero modes with a point defect in two dimensions. For simplicity we consider the 2D square lattice. The sites of the 2D square lattice can be labelled by the $x$ and $y$ coordinates $(i,j)$. The simpliest Majoana hopping model on 2D square lattice with only nearest neighbour hopping has the following Hamiltonian
\begin{equation}
\label{hamiltonian2d}
\mathcal{H}_{2D}=\sum_{(i,j)}it_{i,j}^{x}\eta_{i,j}\eta_{i+1,j}+it_{i,j}^{y}\eta_{i,j}\eta_{i,j+1},
\end{equation}
in which $t_{ij}^{x}$ and $t_{ij}^{y}$ are real numbers. The sites of the model can be divided into $A$ and $B$ sublattices and the hopping paths in the Hamiltonian (\ref{hamiltonian2d}) only connect $A$ sublattice sites to $B$ sublattice sites. The model (\ref{hamiltonian2d}) is thus a simple model and we can have separate zero modes for $A$ and $B$ sublattices $\zeta^{A}$ and $\zeta^{B}$. The possible Majorana zero mode $\zeta^{A,B}=\sum_{i,j}\lambda_{i,j}\eta_{i,j}$ is determined by the commutation relation $\bigg[\sum_{i,j}\lambda_{i,j}\eta_{i,j},\mathcal{H}_{2D}\bigg]=0$. This leads to the following {\it generalized recurrence relation for 2D}
\begin{equation}
\label{recurrence2d}
\lambda_{k-1,l}t_{k-1,l}^{x}+\lambda_{k,l-1}t_{k,l-1}^{y}-\lambda_{k+1,l}t_{k,l}^{x}-\lambda_{k,l+1}t_{k,l}^{y}=0.
\end{equation}
 Furthermore, the recurrence relation (\ref{recurrence2d}) is satisfied by a sufficient but not necessary condition for $x$ direction and $y$ direction separately, $\lambda_{k-1,l}t_{k-1,l}^{x}-\lambda_{k+1,l}t_{k,l}^{x}=0$ and $\lambda_{k,l-1}t_{k,l-1}^{y}-\lambda_{k,l+1}t_{k,l}^{y}=0$. In general 2D Majorana hopping models with zero modes are harder to construct than 1D models. Here we propose and study a simple 2D model whose low energy physics is the $p_{x}+ip_{y}$ topological superconductor.

To introduce the model, we take the unit cell to be two unit squares of the square lattice with dimension $1\times 2$. The hopping coefficients are defined as $A_{1}$ (from point $(i,j)$ to $(i+1,j)$), $A_{2}$ (from $(i,j)$ to $(i,j+1)$), $A_{3}$ (from $(i,j+1)$ to $(i+1,j+1)$) and $A_{4}$ (from $(i,j+1)$ to $(i,j+2)$). In one unit cell, the Majorana fermion on $(i,j)$ and $(i,j+1)$ are labelled as $\eta^{1}$ and $\eta^{2}$ respectively. The Hamiltonian can then be written as
\begin{equation}
\label{Hlatticemodel}
\mathcal{H}=\sum_{\boldsymbol{x}}iA_{1}\eta_{\boldsymbol{x}}^{1}\eta_{\boldsymbol{x}+\boldsymbol{e}_{1}}^{1}+iA_{2}\eta_{\boldsymbol{x}}^{1}\eta_{\boldsymbol{x}}^{2}+iA_{3}\eta_{\boldsymbol{x}}^{2}\eta_{\boldsymbol{x}+\boldsymbol{e}_{1}}^{2}+iA_{4}\eta_{\boldsymbol{x}}^{2}\eta_{\boldsymbol{x}+\boldsymbol{e}_{2}}^{1}.
\end{equation}
In the model we define lattice vectors $\boldsymbol{e}_{1}=(1,0)$, and $\boldsymbol{e}_{2}=(0,2)$. 

In order to see that this model can represent $p_{x}+ip_{y}$ topological superconductor in complex fermions, we pair up the two Majorana fermions in every unit cell and define complex fermion $c_{\boldsymbol{x}}^{\dagger}=\frac{1}{2}(\eta_{\boldsymbol{x}}^{1}+i\eta_{\boldsymbol{x}}^{2})$. In terms of the complex fermion, the Hamiltonian (\ref{Hlatticemodel}) can be written as
\begin{eqnarray}
\begin{aligned}
\label{Hlatticemodel2}
\mathcal{H}=\sum_{\boldsymbol{x}}&iA_{1}(c_{\boldsymbol{x}}+c_{\boldsymbol{x}}^{\dagger})(c_{\boldsymbol{x}+\boldsymbol{e}_{1}}+c_{\boldsymbol{x}+\boldsymbol{e}_{1}}^{\dagger})\\&-A_{2}(c_{\boldsymbol{x}}+c_{\boldsymbol{x}}^{\dagger})(c_{\boldsymbol{x}}-c_{\boldsymbol{x}}^{\dagger})\\&-iA_{3}(c_{\boldsymbol{x}}-c_{\boldsymbol{x}}^{\dagger})(c_{\boldsymbol{x}+\boldsymbol{e}_{1}}-c_{\boldsymbol{x}+\boldsymbol{e}_{1}}^{\dagger})\\&-A_{4}(c_{\boldsymbol{x}}-c_{\boldsymbol{x}}^{\dagger})(c_{\boldsymbol{x}+\boldsymbol{e}_{2}}+c_{\boldsymbol{x}+\boldsymbol{e}_{2}}^{\dagger}).
\end{aligned}
\end{eqnarray}
For the moment we consider the model without defect, namely the coefficients $A_{1}$ to $A_{4}$ are constants for the whole lattice. This allows us to perform Fourier transformation on the complex fermion $c_{\boldsymbol{x}}=\frac{1}{\sqrt{N}}\sum_{\boldsymbol{k}}c_{\boldsymbol{k}}e^{-i\boldsymbol{k}\cdot\boldsymbol{x}}$. After the antisymmetrization, the Hamiltonian (\ref{Hlatticemodel2}) is transformed into the following 
\begin{equation}
\mathcal{H}=\sum_{\boldsymbol{k}}\left(\begin{array}{cc}
c_{\boldsymbol{k}}^{\dagger}&c_{-\boldsymbol{k}}
\end{array}\right)\left(\begin{array}{cc}
A_{\boldsymbol{k}}&B_{\boldsymbol{k}}\\
B_{\boldsymbol{k}}^{*}&-A_{-\boldsymbol{k}}
\end{array}\right)\left(\begin{array}{c}
c_{\boldsymbol{k}}\\c_{-\boldsymbol{k}}^{\dagger}
\end{array}\right),
\end{equation}
in which we define two symbols
\begin{eqnarray}
\begin{aligned}
\label{definitionofAandB}
&A_{\boldsymbol{k}}=(A_{1}+A_{3})\sin(\boldsymbol{k}\cdot\boldsymbol{e}_{1})-A_{2}+A_{4}\cos(\boldsymbol{k}\cdot\boldsymbol{e}_{2}),\\ &B_{\boldsymbol{k}}=(A_{1}-A_{3})\sin(\boldsymbol{k}\cdot\boldsymbol{e}_{1})-iA_{4}\sin(\boldsymbol{k}\cdot\boldsymbol{e}_{2}).
\end{aligned}
\end{eqnarray}

Now we consider the following choice of the coefficients
\begin{equation}
A_{1}=A_{4}=A_{0}, \qquad A_{3}=-A_{0}.
\end{equation}
Under such choice, the two parameters are given by $A_{\boldsymbol{k}}=-A_{2}+A_{0}\cos(\boldsymbol{k}\cdot\boldsymbol{e}_{2})$, and $B_{\boldsymbol{k}}=2A_{0}\sin(\boldsymbol{k}\cdot\boldsymbol{e}_{1})-iA_{0}\sin(\boldsymbol{k}\cdot\boldsymbol{e}_{2})$. By diagonalizing the Hamiltonian we find that when $A_{0}A_{2}>0$, the lowest point of the band is at $|\boldsymbol{k}|=0$. In this situation, when the energy scale is low, one can expand the parameters within the vicinity of $|\boldsymbol{k}|=0$, 
\begin{eqnarray}
\begin{aligned}
&A_{\boldsymbol{k}}\rightarrow (A_{0}-A_{2})-\frac{1}{2}A_{0}(\boldsymbol{k}\cdot\boldsymbol{e}_{2})^{2},\\ &B_{\boldsymbol{k}}\rightarrow 2A_{0}[(\boldsymbol{k}\cdot\hat{\boldsymbol{e}}_{1})-i(\boldsymbol{k}\cdot\hat{\boldsymbol{e}}_{2})],
\end{aligned}
\end{eqnarray}
in which we have used the fact that $|\boldsymbol{e}_{2}|=2|\boldsymbol{e}_{1}|=2$ and $\hat{\boldsymbol{e}}_{2}$ denotes the unit vector along $\boldsymbol{e}_{2}$. This means that the low energy limit of the model is the $p_{x}+ip_{y}$ topological superconductor with pairing field $\Delta=2A_{0}$ being real. Furthermore the parameter $(A_{0}-A_{2})$ plays the role of the chemical potential and the model is topological when $A_{0}-A_{2}>0$.

Next we consider the vortex defect in this model. To this end we observe that in the $p_{x}+ip_{y}$ topological superconductor, the phase of the order parameter corresponds to a rotation of the lattice frame. In particular, the pairing coefficient $\Delta(k_{x}-ik_{y})$ can be written as $|\Delta|e^{i\theta}(k_{x}-ik_{y})$, which is also
\begin{equation}
|\Delta|\bigg[(\cos\theta k_{x}+\sin\theta k_{y})+i(\sin\theta k_{x}-\cos\theta k_{y})\bigg].
\end{equation}
The phase of the order parameter $\Delta$ can thus be intepreted as the following rotation in local coordinate frame,
\begin{equation}
\left(\begin{array}{c}
k_{x}\\k_{y}
\end{array}\right)\rightarrow \left(\begin{array}{cc}
\cos\theta &\sin\theta\\
-\sin\theta&\cos\theta
\end{array}\right)\left(\begin{array}{c}
k_{x}\\k_{y}
\end{array}\right)
\end{equation}
From (\ref{definitionofAandB}) we see that in our model, a vortex can be created by rotating $\hat{\boldsymbol{e}}_{1}$ and $\hat{\boldsymbol{e}}_{2}$ frame around some lattice point. In commensurate with the square lattice structure, the rotation angle can only be multiples of $\frac{\pi}{2}$. For spinless fermion, a full vortex corresponds to $2\pi$ rotation of the frame. This can be created by dividing the lattice into four regions around a certain lattice point. In each of the region, the lattice vectors $\hat{\boldsymbol{e}}_{1}$ are given by $(1,0)$, $(0,1)$, $(-1,0)$ and $(0,-1)$ respectively. Such a lattice structure with vortex is constructed as in Fig. \ref{fig2d}. In Fig. \ref{fig2d}, the bonds with hopping coefficients $A_{2}$ are denoted by thinner lines, as indicated by the topological phase condition $A_{2}-A_{0}<0$; the boundaries between the four regions are given by the blue dashed lines. The sites in the square lattice can be divided into $A$ and $B$ sublattices, we call the sublattice containing the center of the four region (the green dot) as $A$ sublattice. This construction is a lattice realization of the vortex in the continuous field theory of the $p_{x}+ip_{y}$ topological superconductor in Sec. \ref{seccontinuous}.

There exists a localized Majorana zero mode $\zeta^{A}$ in this construction. The solution of its wavefunction $\{\lambda_{i}^{A}\}$ can be obtained by noting the similarity between the lattice structure in Fig. \ref{fig2d} and the 1D SSH model discussed in Sec. \ref{subsecssh}. In Fig. \ref{fig2d}, we label the sites with a nonzero $\lambda_{i}^{A}$ by colored dots; the color green, blue, red etc. denote the {\it magnitude} of $\lambda_{i}^{A}$. If the ratio $|\frac{A_{2}}{A_{0}}|=\hat{\mu}<1$, then we have $|\frac{\lambda_{\text{blue}}}{\lambda_{\text{green}}}|=|\frac{\lambda_{\text{red}}}{\lambda_{\text{blue}}}|=\cdots=\hat{\mu}$. It can be easily checked that such a solution satisfies $[\zeta^{A},\mathcal{H}]=0$. The similarity between the solution of zero mode in this 2D model and the zero mode in the SSH model agrees with the {\it defect classification} \cite{teo2010,chiu16} of the topological phases as the {\it codimension} of the point defect in 1D and 2D models are the same.

\section{Using Majorana orthogonal transformations to construct composite models hosting Majorana zero modes} \label{secgeneralization}

Having discussed simple Majorana hopping models that carry Majorana zero modes, we now move on to composite models and generalizations. In particular, we will be using the result from Sec. \ref{subseczeromode} that for a composite model of independent Majorana layers, the zero modes of the model have one-to-one correspondence with the zero modes in each layer. To construct a composite model with one defect Majorana zero mode, we {\it stack} two layers of Majorana hopping models as in Eq. \ref{generaldoublelayermajorana}, one layer has no Majorana zero mode, the other layer has a single defect Majorana zero mode. By {\it Majorana orthogonal transformation} the Hamiltonian of the system is transformed into Eq. \ref{doublelayer}. In this section we will be focusing on building composite models by stacking two {\it rotated Kitaev chain}. After Majorana orthogonal transformation we obtain one-dimensional models that have {\it spinful fermions} and various types of superconducting pairing, both features help these models to be more relevant in real experiments.

We begin by considering two independent layers of rotated Kitaev chains. For the first chain, the Hamiltonian  $\mathcal{H}_{t,\Delta,\mu,\theta}(\gamma_{i}^{\alpha},\gamma_{i}^{\beta})$ is given by Eq. \ref{rotatedkitaevchain}. For the second chain, we use corresponding $\tilde{\gamma}_{i}^{\alpha,\beta}$ to label its Majorana fermions, and it has another set of parameters $\tilde{t}_{i}$, $\tilde{\Delta}_{i}$, $\tilde{\mu}_{i}$ as well as rotation angle $\phi$; its Hamiltonian can be got from (\ref{rotatedkitaevchain}) and it is given by $\tilde{\mathcal{H}}_{\tilde{t},\tilde{\Delta},\tilde{\mu},\phi}(\tilde{\gamma}_{i}^{\alpha},\tilde{\gamma}_{i}^{\beta})$. The Hamiltonian for the entire system is given by $\mathcal{H}_{t,\Delta,\mu,\theta}\oplus\tilde{\mathcal{H}}_{\tilde{t},\tilde{\Delta},\tilde{\mu},\phi}$ which is written in terms of the Majorana fermions. We then perform the Majorana orthogonal transformation and define two complex fermions 
\begin{equation}
c_{i}^{\dagger}=\frac{1}{2}(\gamma_{i}^{\alpha}+i\tilde{\gamma}_{i}^{\alpha}),\qquad d_{i}^{\dagger}=\frac{1}{2}(\gamma_{i}^{\beta}+i\tilde{\gamma}_{i}^{\beta}).
\end{equation}
In terms of these complex fermions, the total Hamiltonian after the Majorana orthogonal transformation is given by
\begin{eqnarray}
\label{tworotatedkitaevh}
\begin{aligned}
\mathcal{H}_{t,\Delta,\mu,\theta}&\oplus\tilde{\mathcal{H}}_{\tilde{t},\tilde{\Delta},\tilde{\mu},\phi}=\\\frac{1}{2}\sum_{i}&(\tilde{\Delta}_{i}\sin2\phi-\Delta_{i}\sin2\theta)i(c_{i}c_{i+1}-d_{i}d_{i+1})+\\&(\Delta_{i}\sin2\theta+\tilde{\Delta}_{i}\sin2\phi)i(d_{i}^{\dagger}d_{i+1}-c_{i}^{\dagger}c_{i+1})+\\&(\Delta_{i}\cos2\theta-\tilde{\Delta}_{i}\cos2\phi+t_{i}-\tilde{t}_{i})id_{i}c_{i+1}+\\&(\Delta_{i}\cos2\theta-\tilde{\Delta}_{i}\cos2\phi-t_{i}+\tilde{t}_{i})ic_{i}d_{i+1}+\\&(\Delta_{i}\cos2\theta+\tilde{\Delta}_{i}\cos2\phi+t_{i}+\tilde{t}_{i})id_{i}^{\dagger}c_{i+1}+\\&(\Delta_{i}\cos2\theta+\tilde{\Delta}_{i}\cos2\phi-t_{i}-\tilde{t}_{i})ic_{i}^{\dagger}d_{i+1}+\\&(\mu_{i}-\tilde{\mu}_{i})id_{i}c_{i}+(\mu_{i}+\tilde{\mu}_{i})id_{i}^{\dagger}c_{i}+\text{h.c.}.
\end{aligned}
\end{eqnarray}
As can be seen from this general Hamiltonian, different choices of parameters will result in different form of complex-fermion Hamiltonian. Considering possible realizations in experiments, the complex-fermion Hamiltonian should be as simple as possible. In the following we consider three possible sets of parameters of (\ref{tworotatedkitaevh}), all of these choices result in a rather simple spinful Hamiltonian with various types of superconducting pairing.

\subsection{Model one with $\tilde{t}=t$, $\tilde{\Delta}=\Delta$ and $\theta=\phi=0$} \label{subsecmodelone}

For the first case we consider the following choice of parameter for the total Hamiltonian (\ref{tworotatedkitaevh}), $\tilde{t}=t$, $\tilde{\Delta}=\Delta$, $\theta=\phi=0$, which are all constants. The free parameters in the model are the chemical potentials $\mu_{i}$ and $\tilde{\mu}_{i}$. These parameters corresponds to stacking two original Kitaev chains with real pairing given by Eq. \ref{kitaevchainmajoranah}. In this case the total Hamiltonian (\ref{tworotatedkitaevh}) is written as
\begin{eqnarray}
\label{doublekitaevone}
\begin{aligned}
\mathcal{H}\oplus\tilde{\mathcal{H}}=\sum_{i}&(t+\Delta)id_{i}^{\dagger}c_{i+1}+(\Delta-t)ic_{i}^{\dagger}d_{i+1}+\\&\frac{1}{2}(\mu_{i}-\tilde{\mu}_{i})id_{i}c_{i}+\frac{1}{2}(\mu_{i}+\tilde{\mu}_{i})id_{i}^{\dagger}c_{i}+\text{h.c.}.
\end{aligned}
\end{eqnarray}
Now one can use the results for individual Kitaev chain to creat Majorana zero mode in this model. Specifically we set the first chain to be a Kitaev chain with a $\mu$ defect and a corresponding Majorana zero mode and the second chain to be a Kitaev chain without any Majorana zero mode. To achieve this, we take the following choice for the chemical potentials. For $i>0$, $\mu_{i}$ and $\tilde{\mu}_{i}$ both take constant values with $\mu_{i}=\mu_{+}$, $\tilde{\mu}_{i}=\tilde{\mu}_{+}$; and for $i<0$, both $\mu_{i}$ and $\tilde{\mu}_{i}$ are constants different from those for $i>0$, $\mu_{i}=\mu_{-}$, $\tilde{\mu}_{i}=\tilde{\mu}_{-}$. We require that these constants satisfy the relations $\tilde{\mu}_{-}=\mu_{-}$ and $\tilde{\mu}_{+}=2\mu_{-}-\mu_{+}$, as well as $|\mu_{-}|>2|t|$ and $|\mu_{+}|<2|t|$. Since both $|\tilde{\mu}_{\pm}|>2|t|$, the second chain is in non-topological phase and thus hosts no Majorana zero mode. In the first Kitaev chain, there is a $\mu$ defect at $i=0$ separating a topological half and a non-topological half. Therefore the double-layer system (\ref{doublekitaevone}) has one Majorana zero mode at defect $i=0$. Further Majorana orthogonal transformation can simplify the double-layer Hamiltonian, specifically under phase rotation $d_{i}^{\dagger}\rightarrow -id_{i}^{\dagger}$, the Hamiltonian (\ref{doublekitaevone}) can be written as
\begin{eqnarray}
\begin{aligned}
\label{doublelayerkitaevchain2}
\mathcal{H}\oplus\tilde{\mathcal{H}}=\sum_{i}&(t+\Delta)(d_{i}^{\dagger}c_{i+1}+c_{i+1}^{\dagger}d_{i})\\&+(t-\Delta)(c_{i}^{\dagger}d_{i+1}+d_{i+1}^{\dagger}c_{i})\\&+\mu_{-}(d_{i}^{\dagger}c_{i}+c_{i}^{\dagger}d_{i})+\delta_{i}(d_{i}^{\dagger}c_{i}^{\dagger}+c_{i}d_{i}),
\end{aligned}
\end{eqnarray}
in which for $i<0$, $\delta_{i}=0$ and the system is an insulator whereas for $i>0$, $\delta_{i}=\mu_{+}-\mu_{-}$ and the system is a superconductor. So the system (\ref{doublelayerkitaevchain2}) is a superconductor-insulator heterostructure, with a Majorana zero mode at the defect $i=0$. 

In order to make contact with real systems, one can assign spin to the two types of fermions $c_{i}$ and $d_{i}$. Here we choose the following 
\begin{eqnarray}
\begin{aligned}
&d_{2k}\rightarrow c_{2k,\uparrow},\qquad c_{2k+1}\rightarrow c_{2k+1,\uparrow};\\&c_{2k}\rightarrow c_{2k,\downarrow},\qquad d_{2k+1}\rightarrow c_{2k+1,\downarrow}.
\end{aligned}
\end{eqnarray}
With the spin assignment, the Hamiltonian (\ref{doublelayerkitaevchain2}) can be written as $\mathcal{H}\oplus\tilde{\mathcal{H}}=\mathcal{H}_{0}+\mathcal{H}_{soc}+\mathcal{H}_{sc}$, in which
\begin{eqnarray}
\begin{aligned}
\mathcal{H}_{0}=&\sum_{i}\bigg[\sum_{\sigma}t(c_{i,\sigma}^{\dagger}c_{i+1,\sigma})+\mu_{-}c_{i,\uparrow}^{\dagger}c_{i,\downarrow}\bigg]+\text{h.c.},\\
\mathcal{H}_{soc}=&\sum_{k}\bigg[\sum_{\sigma}(\Delta\cdot\sigma)(c_{2k,\sigma}^{\dagger}c_{2k+1,\sigma}-c_{2k+1,\sigma}^{\dagger}c_{2k+2,\sigma})\bigg]\\&+\text{h.c.},\\
\mathcal{H}_{sc}=&\sum_{k}\bigg[\delta_{k}(c_{2k,\uparrow}^{\dagger}c_{2k,\downarrow}^{\dagger}-c_{2k+1,\uparrow}^{\dagger}c_{2k+1,\downarrow}^{\dagger})\bigg]+\text{h.c.}.
\end{aligned}
\end{eqnarray}
In the spin-orbit coupling term $\mathcal{H}_{soc}$, we assume $\Delta\cdot\sigma=\Delta$ for $\sigma=\uparrow$ and $\Delta\cdot\sigma=-\Delta$ for $\sigma=\downarrow$. It is noteworthy that further phase rotations may bring the Hamiltonian into simplier form, for example, $c_{2k+1,\sigma}^{\dagger}\rightarrow ic_{2k+1,\sigma}^{\dagger}$.

\subsection{Model two with $\tilde{t}=-t$, $\tilde{\Delta}=\Delta$, $\theta=\phi=\frac{\pi}{4}$ and the Creutz Majorana model} \label{subsecmodeltwo}

For the second case we consider the following choice of parameters $\tilde{t}=-t$, $\tilde{\Delta}=\Delta$ and rotation angle $\theta=\phi=\frac{\pi}{4}$ which are all constants. The free parameters in the model are the chemical potentials $\mu_{i}$ and $\tilde{\mu}_{i}$. With these parameters the total Hamiltonian (\ref{tworotatedkitaevh}) is written as
\begin{eqnarray}
\label{modeltwocreutz}
\begin{aligned}
\mathcal{H}\oplus\tilde{\mathcal{H}}=&\sum_{i}i\Delta(d_{i}^{\dagger}d_{i+1}-c_{i}^{\dagger}c_{i+1})+it(d_{i}c_{i+1}-c_{i}d_{i+1})\\&+\frac{1}{2}(\mu_{i}-\tilde{\mu}_{i})id_{i}c_{i}+\frac{1}{2}(\mu_{i}+\tilde{\mu}_{i})id_{i}^{\dagger}c_{i}+\text{h.c.}.
\end{aligned}
\end{eqnarray}
As for model one in the previous section, we can assign spins to the complex fermions to achieve a simplier form for the Hamiltonian. To this end, we make the following definition
\begin{equation}
\label{definitionofspinfulcone}
c_{i}\rightarrow ic_{i\uparrow}^{\dagger}, \qquad d_{i}\rightarrow c_{i\downarrow},
\end{equation}
in which $c_{i\uparrow}$ and $c_{i\downarrow}$ are the two components of a spinful complex fermion. Note that the definition (\ref{definitionofspinfulcone}) involves a particle-hole transformation and an extra phase. 

For simplicity, here we focus on the boundary zero modes and thus assume that $\mu_{i}$ and $\tilde{\mu}_{i}$ are both constants. To further simplify the Hamiltonian we introduce two constants $\omega$ and $\tau$ such that $\mu_{i}=\omega+\tau$ and $\tilde{\mu}_{i}=\tau-\omega$; also for the spinful fermion we introduce the Dirac spinor $\psi_{i}=\left(c_{i\uparrow},c_{i\downarrow}\right)^{T}$ for every site. The Hamiltonian (\ref{modeltwocreutz}) is then written as
\begin{eqnarray}
\label{creutzmajoranamodel}
\begin{aligned}
\mathcal{H}\oplus\tilde{\mathcal{H}}=\sum_{i}& \psi_{i}^{\dagger}(t\sigma^{x}-i\Delta\sigma^{z})\psi_{i+1}+\frac{1}{2}\omega\psi_{i}^{\dagger}\sigma^{x}\psi_{i}\\&+\tau c_{i\uparrow}^{\dagger}c_{i\downarrow}^{\dagger}+\text{h.c.},
\end{aligned}
\end{eqnarray}
in which $\sigma^{x}$ and $\sigma^{z}$ are Pauli matrices. This model (\ref{creutzmajoranamodel}) is also known as the {\it Creutz Majorana model} \cite{sticlet2014,Creutz94,Piga17}.

For the existence of a single Majorana boundary mode, we require that one and only one of the two rotated Kitaev chain is in topological phase. This means that we either have $|\omega+\tau|<2|t|$ or $|\tau-\omega|<2|t|$. To this end we can define a $Z_{2}$ topological number for the model, 
\begin{equation}
\mathcal{M}=\operatorname{sgn}[(\omega+\tau+2t)(\omega+\tau-2t)(\omega-\tau+2t)(\omega-\tau-2t)]. 
\end{equation}
If one of the layer is in the topological phase the number is $-1$, and we have single Majorana boundary mode; otherwise the number is $+1$ and we have either zero Majorana mode or two Majorana modes on the boundary, these two situations are equivalent in the sense that the two Majorana boundary modes are not stable against local perturbations. The topological number $\mathcal{M}$ agrees with the {\it Majorana number} obtained by Ref. \onlinecite{sticlet2014} from analysis of the spectrum in momentum space. Following the discussion in Sec. \ref{subsecmodelone} the defect zero modes of the model (\ref{creutzmajoranamodel}) for an infinite chain can be discussed in a similar way.

\subsection{Model three with $\tilde{\mu}=\mu$, $\tilde{\Delta}=\Delta$ and $\theta=\phi=\frac{\pi}{4}$}

For the third case we consider the following choice of parameter $\tilde{\mu}=\mu$, $\tilde{\Delta}=\Delta$, $\theta=\phi=\frac{\pi}{4}$, which are all constants. The free parameters in this case are the hopping coefficients $t_{i}$ and $\tilde{t}_{i}$. The total Hamiltonian (\ref{tworotatedkitaevh}) can thus be written as
\begin{eqnarray}
\label{modelthreeh}
\begin{aligned}
\mathcal{H}\oplus\tilde{\mathcal{H}}=\sum_{i}&i\Delta(d_{i}^{\dagger}d_{i+1}-c_{i}^{\dagger}c_{i+1})\\&+\frac{1}{2}(t_{i}-\tilde{t}_{i})i(d_{i}c_{i+1}-c_{i}d_{i+1})\\&+\frac{1}{2}(t_{i}+\tilde{t}_{i})i(d_{i}^{\dagger}c_{i+1}-c_{i}^{\dagger}d_{i+1})\\&+\mu id_{i}^{\dagger}c_{i}+\text{h.c.}.
\end{aligned}
\end{eqnarray}
As with model one and model two, we can assign spins to the complex fermions. In this model, we choose the following
\begin{equation}
d_{i}\rightarrow c_{i\uparrow}, \qquad c_{i}\rightarrow -ic_{i\downarrow}
\end{equation}
And we define the Dirac spinor for the spinful complex fermion on every site, $\psi_{i}=\left(c_{i\uparrow},c_{i\downarrow}\right)^{T}$. Similarly with model two in Sec. \ref{subsecmodeltwo} we define the hopping coefficients as $t_{i}=\omega+\tau$ and $\tilde{t}_{i}=\omega-\tau$ to simplify the notations. Then the total Hamiltonian (\ref{modelthreeh}) is brought into the following form,
\begin{eqnarray}
\label{modelthreehamiltonian}
\begin{aligned}
\mathcal{H}\oplus\tilde{\mathcal{H}}=\sum_{i}&\psi_{i}^{\dagger}\big(i\Delta\sigma^{z}+\omega\sigma^{x}\big)\psi_{i+1}+\frac{1}{2}\mu \psi_{i}^{\dagger}\sigma^{x}\psi_{i}\\&+\tau\big(c_{i\uparrow}c_{i+1,\downarrow}+c_{i+1,\uparrow}c_{i\downarrow}\big)+\text{h.c.},
\end{aligned}
\end{eqnarray}
in which $\sigma^{x}$ and $\sigma^{z}$ are Pauli matrices. 

The form of the Hamiltonian (\ref{modelthreehamiltonian}) is different from the second model (\ref{creutzmajoranamodel}) in that the pairing term is between fermions with opposite spin on the {\it neighbouring sites}. For the existence of boundary Majorana zero mode, we again require that one and only one of the rotated Kitaev chain is in topological phase. That means we either have $|\mu|<2|\omega+\tau|$ or $|\mu|<2|\omega-\tau|$. To this end we can define the following $Z_{2}$ topological number,
\begin{equation}
\mathcal{M}=\operatorname{sgn}[(\omega+\tau+\frac{\mu}{2})(\omega+\tau-\frac{\mu}{2})(\omega-\tau+\frac{\mu}{2})(\omega-\tau-\frac{\mu}{2})]
\end{equation}
If the number is $-1$ then we have a single Majorana zero mode on the boundary of the system. The defect zero mode for an infinite chain in this model can be discussed in a similar way as model one.

\subsection{Further applications}

The building blocks of the composite lattice models considered in this work are limited to {\it simple models} or those which can be transformed into simple models by Majorana orthogonal transformations. Certainly models which cannot be transformed into simple models can be introduced. For example, the Kitaev chain parameters $t$, and $\Delta$ in (\ref{kitaevchainhamiltonian}) can be chosen in a way that are not possible to be transformed into real numbers simultaneously by phase rotations. The possible zero modes on the edges and defects in this type of model need separate considerations. 

On a broader perspective, the method of Majorana orthogonal transformation can be used in any 1D lattice models; it gives a {\it real-space} perspective independent of momentum-space spectrum analysis for existing models which are not limited to the ones considered here \cite{Wakatsuki14,zhao2014}; and it can be used to construct new models hosting Majorana zero modes. Another possible generalization of the method is in exactly solvable interacting models, including the ones discussed in Sec. \ref{subsecexactlysolvable} and in spin chains. Specifically, some spin chains can be mapped into {\it free} fermionic models by the Jordan-Wigner (JW) transformation \cite{fradkinbook,jordan28,lieb61} and thus have corresponding topological properties of the free fermionic models \cite{niu2012}. 

The methods can be applied to 2D lattice models as well. In particular we have constructed a model in Sec. \ref{subsec2dlattice} which has point-defect Majorana zero modes. One can introduce another layer of Majorana fermions with identical lattice structure but hopping coefficients all equal on the bonds. The new layer host no localized Majorana zero mode, therefore the two layer system has a single defect Majorana zero mode. However, the system Hamiltonian after Majorana orthogonal transformation is complex with different superconducting pairing coefficients and hopping coefficients on neighbouring bonds, we will not discuss it in detail here.

\section{Conclusion and outlook} \label{secconclusion}

In this paper we have discussed the application of Majorana orthogonal transformation to the study of Majorana zero modes in various models. Specifically for the 2D continuous $p_{x}+ip_{y}$ topological superconductor, we show that the doubled system can be described by a U(1) topological gauge theory of massive Dirac fermion, using which one can compute the degeneracy splitting of multiple vortex Majorana zero modes. For lattice models hosting Majorana zero modes, a {\it real space analysis} on the wavefunction of the Majorana zero modes is performed. To this end, we introduce the concept of {\it simple models} and the decoupled SSH model and the Kitaev chain are considered as examples. We then discuss generalizations of simple models by considering the rotated Kitaev chain. For 2D lattice models, we construct a simple Majorana layer that represents the $p_{x}+ip_{y}$ superconductor at low energy which hosts defect Majorana zero mode. Finally we show that Majorana orthogonal transformation can be used to construct {\it composite models} with Majorana zero modes. To do this, one stacks two layers of the models, each of which has its own Majorana zero mode distribution. The Majorana orthogonal transformation then glue the two layers together into a single model which {\it inherits} the independent zero modes of the two layers. To this end, three examples are discussed from stacking two rotated Kitaev chains. These composite models can have spinful complex fermion as matter field and are thus more relevant to experiments.

The construction of composite lattice models hosting Majorana zero modes and the doubling of continuous model by Majorana orthogonal transformation indicate that {\it free fermionic topological systems can be added together} to form another topological system. Indeed, this has been used to argue that equivalent classes of defect Hamiltonians have {\it group structure} \cite{teo2010,chiu16,kitaev09} and our construction gives real space examples for this process. With regard to topological phases, our real space analysis does not rely on the topological numbers computed in the momentum space; but it may have some relations with the topological numbers computed in real space \cite{Kitaev2006,bianco11,prodan10,prodan11}, such relations are left for future study. Another natural question to ask is how the different symmetries in complex fermion models manifest in the Majorana hopping models. In particular, different pairings of the Majorana fermions may result in different time-reversal properties of each Majorana fermion and thus may lead to different physical symmetries of the resulting complex fermion models, but they will not change the topological properties such as the existence of zero modes. Detailed analysis on this is left for the future.

Our discussion on the Majorana zero modes has been restricted to models without disorders. But it is easy to see that the defect Majorana zero mode in decoupled SSH model is robust under disorders. The effects of disorder in the Kitaev chain have also been studied \cite{hegde2016}. To study the effects of disorders on the Majorana zero modes, the methods based on the number sequence of the wavefunction applied in this paper can be generalized to {\it transfer matrix method} \cite{hegde2016,degottardi20131,Degottardi132,padavi20181}. Another direction that is worth exploring is the effects of interaction. In general it is expected that interaction has a strong influence on topological phases \cite{morimoto2015,yao2013}. Specifically for 1D system, Fidkowski and Kitaev considered stacking a number of Kitaev chain together and showed that {\it with time reversal symmetry}, the interaction terms breaks the original $Z$ topological classification of the Kitaev chain down to $Z_{8}$ \cite{fidkowski2010,fidkowski2011}. The influence of interaction terms on the Majorana zero modes in 1D has been studied for different systems \cite{katsura2015,gergs2016,mcginley2017,Marques17}. From our perspective, it is possible to discuss the Majorana orthogonal transformations for the interacting systems with quartic Majorana terms. These topics are left for future study.

\section*{Acknowledgements}

The author thanks G. Chen and C. Wang for useful discussions. This work is supported by Research Grants Council of Hong Kong with General Research Fund Grant No.17303819.


\bibliography{refJFU}

\end{document}